\documentclass[]{article}

\pdfoutput=1
\usepackage{authblk}
\usepackage{fullpage}
\usepackage{amsmath}
\usepackage{booktabs}

\usepackage{caption}
\usepackage{graphicx}

\usepackage[hidelinks]{hyperref}
\usepackage[round, semicolon]{natbib}
\usepackage{multirow}
\usepackage{lscape}

\title{Mapping Depth to Bedrock, Shear Stiffness, and Fundamental Site Period at CentrePort, Wellington using Surface Wave Methods: Implications for Local Seismic Site Amplification}
\author[1]{Joseph P. Vantassel}
\date{}
\author[1]{Brady R. Cox}
\author[2]{Liam Wotherspoon}
\author[1]{Andrew Stolte}
\affil[1]{The University of Texas at Austin}
\affil[2]{University of Auckland}

\begin{document}

\maketitle

\begin{abstract}
Wellington's port (CentrePort) experienced significant damage from the $M_w$ 7.8 Kaik\=oura earthquake as a result of soil liquefaction, lateral spreading, and shaking-induced damage to structures. To investigate these ill effects, and propose mitigation measures to prevent similar damage in future earthquakes, there was a need to quantify the variations in the depth to bedrock, shear stiffness, and fundamental site period ($T_0$) across the port. In order to characterize $T_0$ and develop shear wave velocity (Vs) profiles for use in seismic site response analyses, horizontal-to-vertical (H/V) spectral ratio measurements and active-source and passive-wavefield surface wave testing (i.e., MASW and MAM, respectively) were performed across the port. A site period map developed from 114 H/V spectral ratio measurements indicates several areas of rapidly changing, complex subsurface structure. Deep (200-plus meters) Vs profiles developed at six reference locations across the port were used to estimate the depth to soft (Vs $>$ 760 m/s) and hard (Vs $>$ 1500 m/s) rock. $T_0$ estimates from H/V spectral ratio measurements ($T_{0,H/V}$) at the six reference locations are shown to be related to the depth of hard rock based on linear viscoelastic transfer functions calculated from Vs profiles truncated at several depths. $T_{0,H/V}$ measurements at two ground motion stations near the port are also shown to be in reasonably good agreement with predominant periods of maximum spectral amplification recorded during both the 2016 Kaik\=oura and 2013 Cook Strait earthquakes, despite these sites also being effected by soil nonlinearity and potential 3D basin edge effects. 
\end{abstract}


\pagebreak

\section*{Introduction}

The $M_w$ 7.8 Kaik\=oura earthquake caused significant damage to the Wellington port (CentrePort) as a result of soil liquefaction, lateral spreading, and shaking-induced damage to structures. The observed liquefaction effects, results from subsequent detailed geotechnical investigations, and preliminary liquefaction analyses at CentrePort are presented in Cubrinovski et al. (\citeyear{cubrinovski_liquefaction-induced_2018}). This paper presents findings from a comprehensive dynamic site characterization study ultimately aimed at understanding the spatially- and frequency-dependent amplification of ground motions experienced on the thick, soft soils (combined reclamation fill and native deposits) beneath CentrePort. As discussed by Bradley et al. (\citeyear{bradley_ground_2017}), ground motions recorded on soft soils throughout Wellington during the Kaik\=oura earthquake had spectral amplifications across a wide range of vibration periods that exceeded the amplification factors prescribed by the NZS1170.5 seismic loading provisions (Standards New Zealand, \citeyear{standards_new_zealand_structural_2004}). While 1D site response analyses can lend insights into some of the factors causing greater-than-expected amplification, the complicated 3D nature of the subsurface beneath Wellington needs to be better characterized so that more rigorous dynamic analyses, including for example basin-edge generated surface waves, can be performed. This paper presents important findings that will inform refinements to the 3D velocity structure beneath a key area of the city.

Semmens et al. (\citeyear{semmens_its_2010}) developed a 3D model of central Wellington to characterize the variation in depth to bedrock and shear stiffness of the overlying soil deposits. However, in the vicinity of CentrePort they were not able to locate any detailed information about the depth to bedrock, nor make any new measurements to infer soil shear stiffness and fundamental site period ($T_0$). Thus, their estimates of the spatial variation in $T_0$ and depth to bedrock beneath the port were based on inferences from an understanding of regional geology and extrapolations from measurements made hundreds of meters from the boundaries of the port. Following the Kaik\=oura earthquake, our team was granted access to the port for the purpose of conducting a non-invasive dynamic site characterization study. Our efforts included ambient vibration horizontal-to-vertical (H/V) spectral ratio measurements at 114 locations and deep (200-plus meters) shear wave velocity (Vs) profiling via combined active-source and passive-wavefield surface wave testing at six reference locations across the port. A detailed fundamental site period map for the port has been developed from the H/V data, and estimates for the depth of bedrock have been made from Vs profiles developed through joint inversion of the surface wave dispersion and H/V data.

To place our measurements in context, we first briefly discuss the geology beneath CentrePort. We then discuss our testing methodologies and results, which clearly illustrate complex subsurface conditions and abrupt changes in fundamental site period and depth to bedrock over relatively short distances at several locations. We also note that estimates of site period from H/V data ($T_{0,H/V}$) are further complicated by azimuthal dependency, particularly in areas where the subsurface is inferred to be particularly erratic and potentially influenced by faulting and/or paleochannels. The measured $T_{0,H/V}$ values near the ground motion stations PIPS and CPLB are then compared with the period range of maximum spectral amplifications observed in the 2016 $M_w$ 7.8 Kaik\=oura and 2013 $M_w$ 6.6 Cook Strait (also referred to as Seddon) earthquakes. For additional information on the Cook Strait earthquake the reader is directed to Holden et al. (\citeyear{holden_sources_2013}).

\section*{Geology of CentrePort}

\begin{figure}[!]
    \centering
	\includegraphics[width=0.6\textwidth]{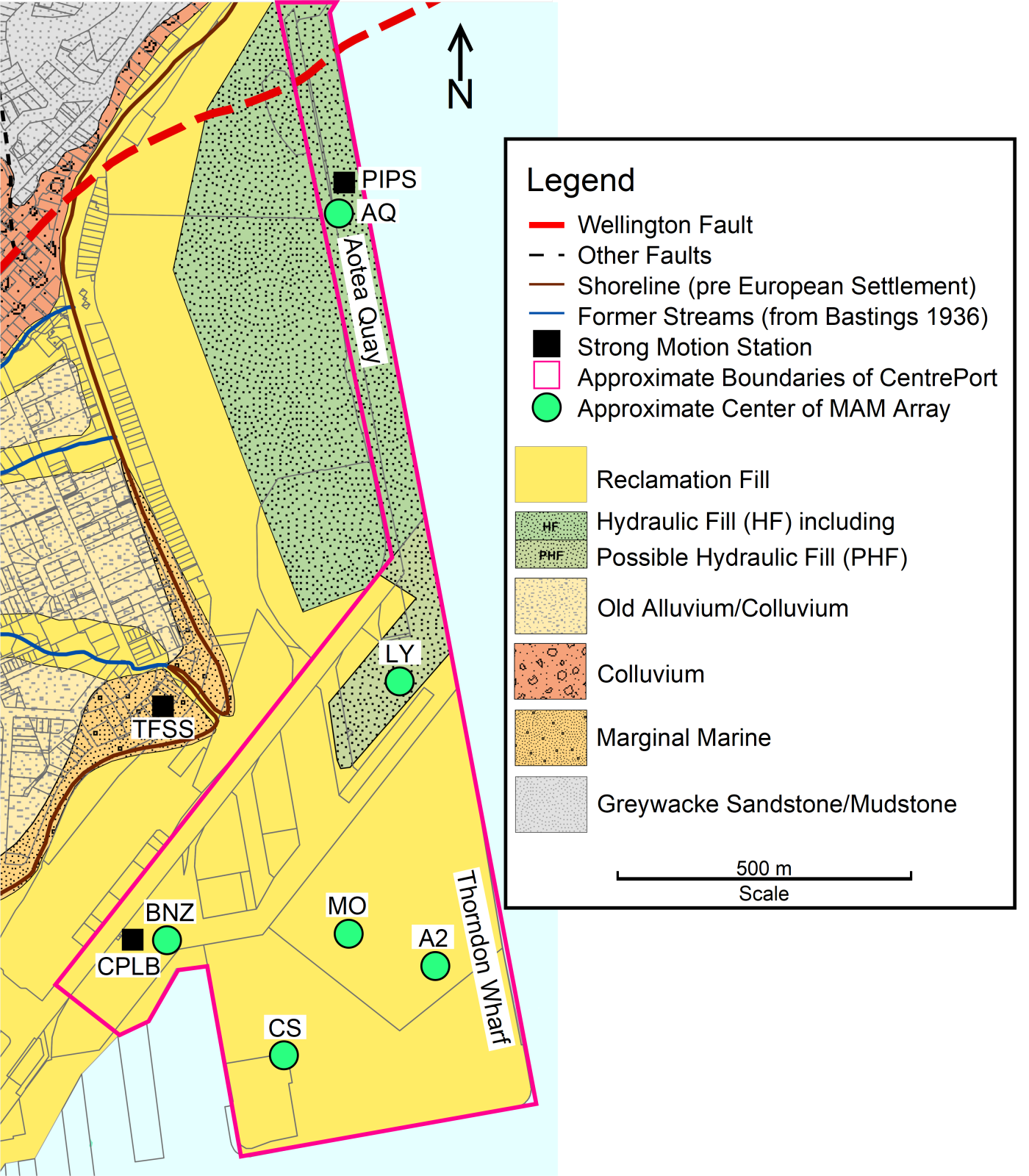}
	\caption{Surface geology of CentrePort illustrating the regions of hydraulic fill and common reclamation fill, also referred to as end-tipped fill (adapted from Semmens et al. \citeyear{semmens_its_2010}). Also shown are the locations of three strong motion stations that recorded the $M_w$ 7.8 Kaik\=oura earthquake and six reference locations where surface wave arrays were used for deep (200-plus meters) Vs profiling.}
	\label{fig:1}
\end{figure}

CentrePort generally resides on 10-20 m of reclaimed soils deposited over 1-5 m of marine sediments, underlain by 100-plus meters of alluvium \citep{cubrinovski_liquefaction-induced_2018}. The reclaimed soils consist of two main types (refer to Figure \ref{fig:1}): hydraulic fill, which consists of sediments dredged from Wellington harbor, and common reclamation fill, also referred to as end-tipped fill, which consists of a gravel-sand-silt mixture (Tonkin \& Taylor Ltd., \citeyear{tonkin__taylor_ltd_thorndon_2012}). The port is roughly divided into two regions based on surficial fill type: the northern region, which is underlain by dredged hydraulic fill, and the southern region, which is underlain by the common reclamation/end-tipped fill. The northern region includes Aotea Quay, the PIPS strong motion station, and two of the surface wave arrays used for dynamic site characterization [Aotea Quay (AQ) and Log Yard (LY)]. The southern region includes the Thorndon Wharf area, the CPLB strong motion station, and four surface wave arrays [Main Office (MO), BNZ Building (BNZ), Cold Store (CS), and Thorndon Wharf (A2)]. Dividing the northern hydraulic fill region from the southern common reclamation fill region is a buried sea wall, located just south of the Log Yard, an artifact of a previous reclamation that runs collinear with the southeast edge of the area labeled in Figure 1 as possible hydraulic fill. For a complete discussion of the construction process, materials used, and sequencing of the reclamations beneath CentrePort the reader is referred to Cubrinovski et al. (\citeyear{cubrinovski_liquefaction-induced_2018}).

The Semmens et al. (\citeyear{semmens_its_2010}) surface geology map shown in Figure 1 also clearly indicates several former river channels/valleys that incise the old alluvium/colluvium deposits west of the pre European Settlement-shoreline (e.g., near TFSS) and presumably extend into the harbor beneath the reclamation fill. As discussed in greater detail below, Semmens et al. (\citeyear{semmens_its_2010}) estimated the depth to greywacke bedrock beneath the port to range from a minimum depth of about 130 m in the vicinity of CPLB to a maximum of about 300 m just north of PIPS, near the Wellington Fault.

\section*{Surface Wave Testing Methods}

\begin{figure}[!]
    \centering
	\includegraphics[width=0.6\textwidth]{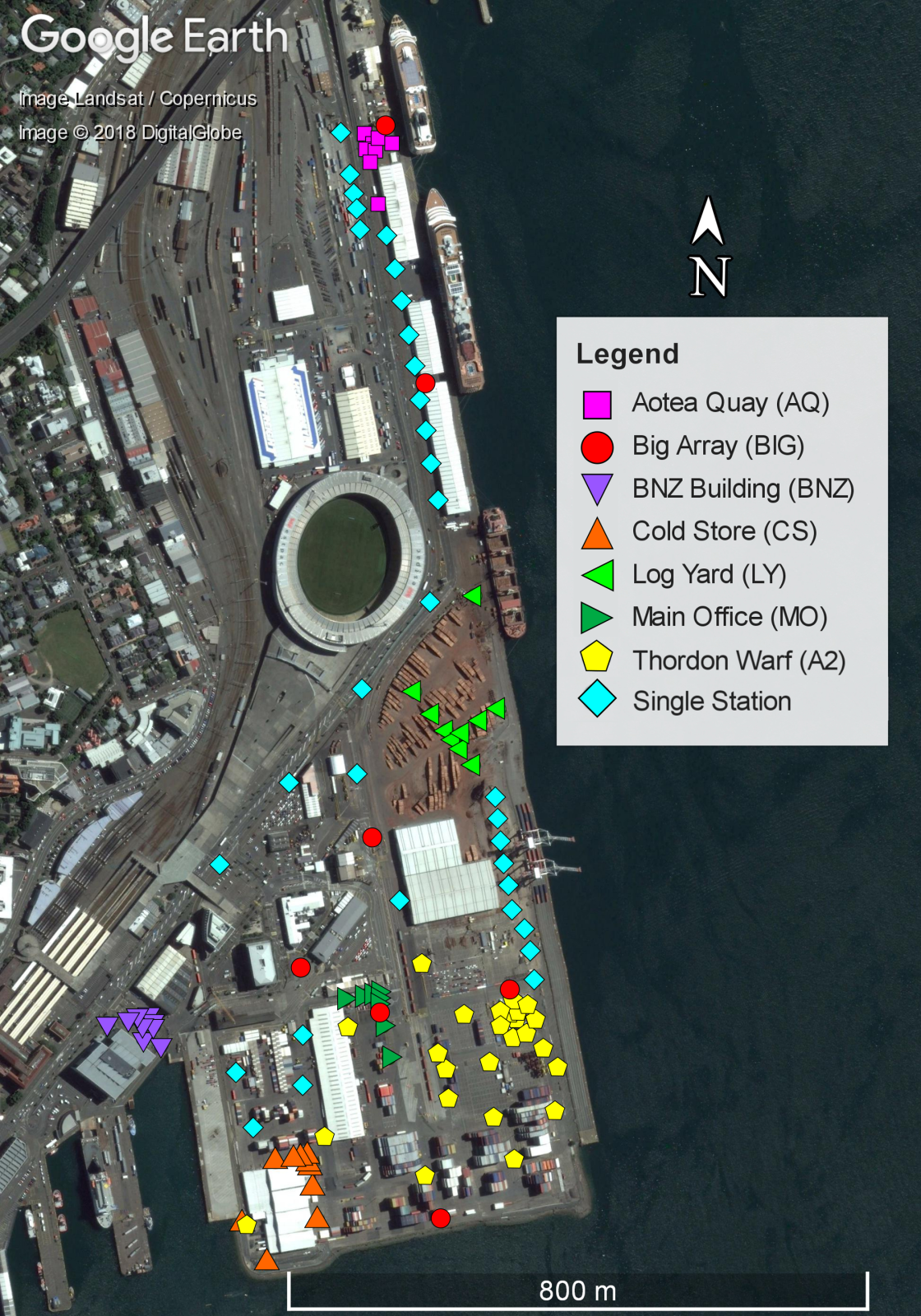}
	\caption{Locations of three-component broadband seismometers used to record ambient vibrations for 2D MAM arrays and single station H/V spectral ratio measurements across CentrePort.}
	\label{fig:2}
\end{figure}

Our team conducted active-source multi-channel analysis of surface waves (MASW) testing and passive-wavefield 2D microtremor array measurements (MAM) at six reference locations across CentrePort in May and June of 2017. The approximate center of each 2D MAM array is shown in Figure 1. The locations of each three-component broadband seismic station comprising the various 2D MAM arrays are shown in Figure \ref{fig:2}. Of the 114 station locations used to record ambient vibrations across the port, 81 were deployed in 2D MAM arrays of various geometries to facilitate extraction of surface wave dispersion data. The remaining 33 station locations, referred to as single station measurements, were used to supplement the MAM arrays by providing additional spatial coverage for H/V measurements across the port.

\section*{Horizontal-to-Vertical Spectral Ratios}

H/V spectral ratios from ambient vibration measurements were calculated for each of the 114 station locations shown in Figure 2. At each location, a three-component Nanometrics, Inc.\ Trillium Compact 20 s seismometer with a flat frequency response between 20 seconds and 100 Hz was used to record ambient vibrations. Seismometers were set on an aluminum baseplate directly on the pavement, leveled, oriented to magnetic north, and shielded with a weighted plastic cover to mitigate effects of wind vibration. The data acquisition system consisted of Nanometrics, Inc.\ Centaur digitizers (24 bit ADC, 135 dB dynamic range). Time records were digitized at a sampling frequency of 100 Hz, with durations ranging from 30 minutes for the single station deployments up to 9 hours for the largest MAM array.

H/V spectral ratios were computed by dividing the continuous, three-component ambient vibration records into individual time windows between 60- and 120-s in length. A cosine taper of 5\% of the window length was utilized on the ends of each window to avoid complications when transforming to the frequency domain. Frequency spectra for each window were smoothed using Konno and Ohmachi (\citeyear{konno_ground-motion_1998}) smoothing with a bandwidth coefficient of 40. After smoothing, spurious windows caused by transient nearfield noise (e.g., vehicles passing near the sensor) were removed based on their anomalous frequency domain response. To represent the two horizontal components as a single component, the squared average (i.e., square root of the average of the squared components) was computed for each frequency. The mean H/V spectral ratio curve for a given station was calculated as the mean of the curves calculated from each time window. Where each time window curve was calculated as the ratio between the squared average horizontal and vertical Fourier amplitudes.

A well-defined, dominant frequency peak in the H/V data ($f_{0,H/V}$) can be used to infer the fundamental shear wave resonant frequency of the site ($f_{0,S}$) \citep{lermo_site_1993, lachet_numerical_1994, sesame_guidelines_2004} and/or the lowest-frequency peak of the fundamental mode Rayleigh wave ellipticity ($f_{0,R}$) \citep{malischewsky_loves_2004, poggi_estimating_2010}. When a strong impedance contrast is present at a site, $f_{0,H/V}$, $f_{0,S}$, and $f_{0,R}$ are approximately equal to one another. When a more moderate impedance contrast is present, $f_{0,H/V}$ may be more representative of $f_{0,S}$ \citep{bonnefoy-claudet_nature_2004}. Therefore, H/V spectral ratio measurements can be used as a tool to rapidly develop estimates of fundamental site period ($T_0$ $\approx$ 1/$f_{0,H/V}$) provided a clear, dominant peak exists (or multiple clear peaks if multiple impedance contrasts are present in the soil profile).

\begin{figure}[!]
    \centering
	\includegraphics[width=0.5\textwidth]{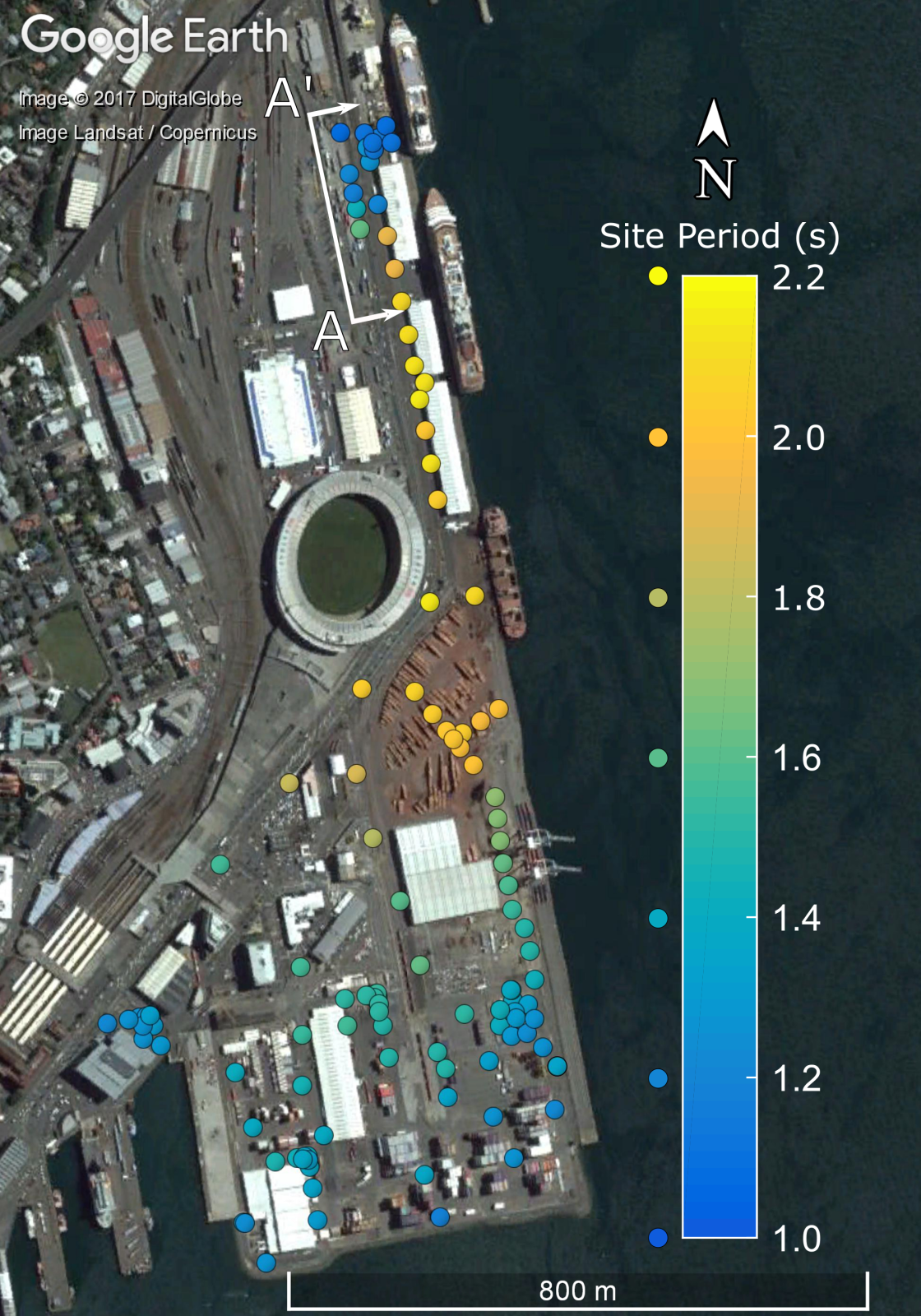}
	\caption{Fundamental site period map inferred from 114 H/V spectral ratio measurements ($T_{0,H/V}$) across CentrePort.}
	\label{fig:3}
\end{figure}

A map showing the variation in fundamental site period across the port inferred from the H/V spectral ratio measurements is shown in Figure \ref{fig:3} decreases rapidly down to approximately 1.0 s for the northernmost stations along Aotea Quay. There are two locations across the port where the site period changes abruptly that deserve further discussion. The first location is in the vicinity of the Log Yard, where the stiffer end-tipped fill of the southern part of the port meets the old sea wall and the softer hydraulic fill of the northern part of the port. In this area $T_{0,H/V}$ changes from approximately 1.7 s to 2.0 s over a distance of about 100 m. The second location of abruptly changing $T_{0,H/V}$ exists along the northern portion of Aotea Quay, where the periods rapidly decrease from approximately 2.1 s to 1.0 s over a distance of approximately 200 m. Causes for these abrupt changes in $T_{0,H/V}$ may be inferred by understanding that the fundamental site period can be approximated using the quarter-wavelength relationship of Equation 1, where it is assumed that a single representative soil layer of thickness H with average shear wave velocity (Vs,avg) overlies a rigid half space.

\begin{equation}
    T_0 \approx 4 \frac{H}{Vs_{avg}}
\end{equation}

Based on Equation 1, it is intuitive that a change in Vs of the overlying soil will have an inverse effect on the fundamental site period (e.g., a decrease in $Vs_{avg}$ would tend to increase $T_0$). However, the fundamental site period is also directly proportional to the thickness of the soil/depth to bedrock (e.g., an increase in H would tend to increase $T_0$). As discussed below, we believe both of these factors are contributing to the rapid increase in $T_{0,H/V}$ near the Log Yard.

\begin{figure}[!]
    \centering
	\includegraphics[width=0.8\textwidth]{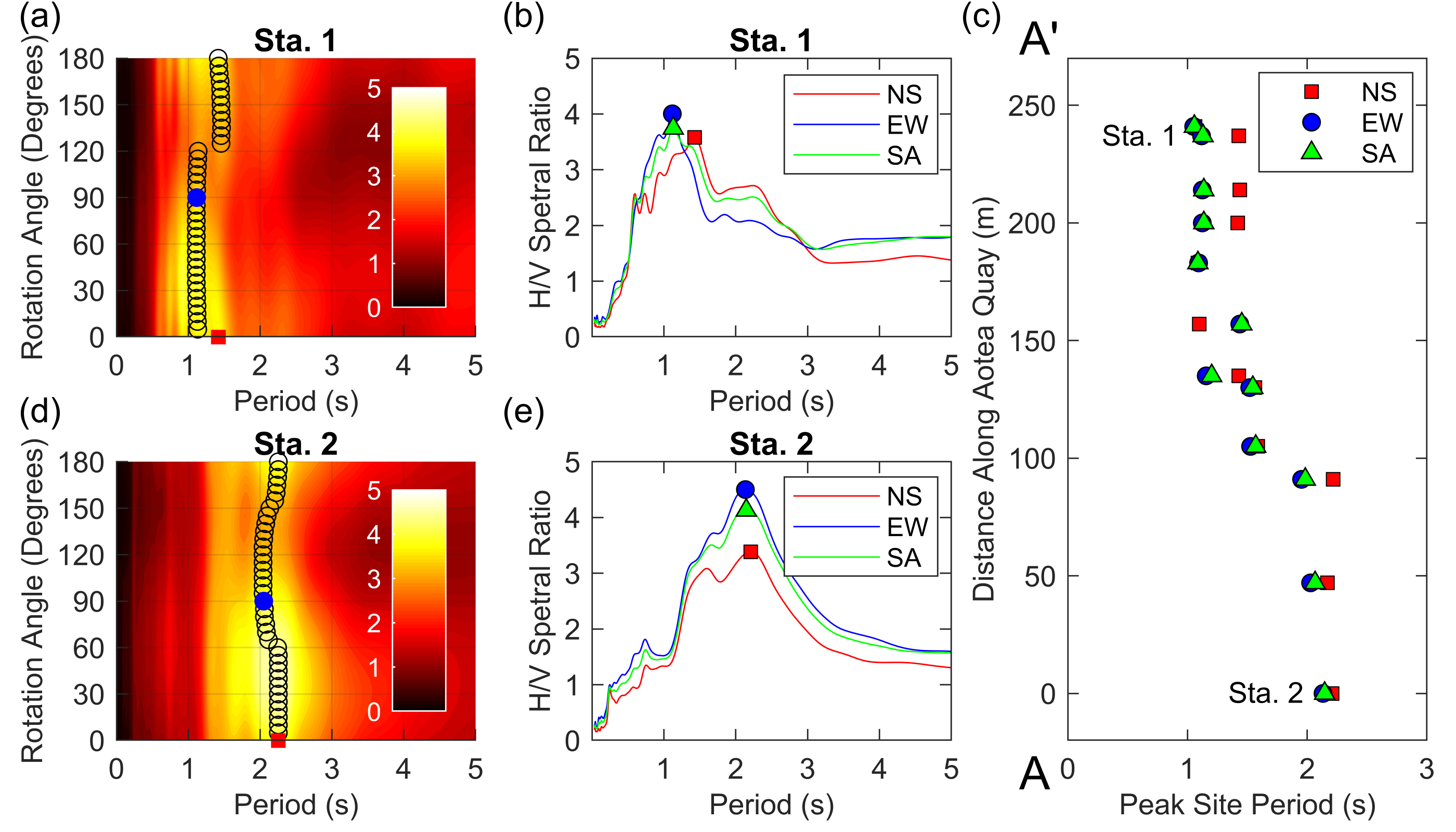}
	\caption{Comparison of H/V spectral ratio measurements along Aotea Quay: (a) example H/V spectral ratio data where $T_{0,H/V}$ changes with azimuth, (b) azimuthal slices/components showing how the period and amplitude of the mean H/V spectral ratio curves for the NS, EW, and SA (squared average) components differ, (c) a comparison of the transition in $T_{0,H/V}$ periods and azimuthal variability for 12 stations along section A-A’, (d) example H/V spectral ratio data where $T_{0,H/V}$ does not change significantly with azimuth, and (e) azimuthal slices/components showing how the period of the mean H/V spectral ratio curves for the NS, EW, and SA (squared average) components agree despite differences in peak amplitude. Note that the location of section A-A' is shown in Figure 3.}
	\label{fig:4}
\end{figure}

The abruptly changing $T_{0,H/V}$ values along the northern portion of Aotea Quay are presented in more detail in Figure \ref{fig:4} by considering a section from A-A' (location indicated in Figure 3). The H/V data shown for Sta.\ 1 in Figure 4a and 4b is considered to be representative of stations located in this zone of rapidly changing site period. The changes in site period in this region are believed to be (at least in part) related to azimuthal dependency. To illustrate this, Figure 4a shows how the fundamental site period changes as the horizontal components are rotated in 5 degree increments. Figure 4b shows two azimuthal slices/components from Figure 4a (i.e., NS [0 degrees] and EW [90 degrees]) in comparison with the SA of these components. Variability in $T_{0,H/V}$ and its amplitude with azimuth were found to be typical for the majority of stations deployed in this area (i.e., northern Aotea Quay). However, a few stations to the south did not show azimuthal variability in site period despite showing variability in amplitude.  Figure 4c, documents the transition in $T_{0,H/V}$ with distance along the quay, as well as illustrates some of the azimuthal variability by indicating $T_{0,H/V}$ for the NS, EW, and SA components. Note how the peak site period is decreasing while the azimuthal variability in site period is increasing along the quay. The H/V data for Sta.\ 2 is presented in Figure 4d and 4e as a typical case for the majority of stations south of the northernmost part of Aotea Quay. These stations showed limited azimuthal variability in site period, but may or may not have shown azimuthal variability in amplitude.

The exact cause of the azimuthal variability remains unclear and requires further research. However, the use of 2D array processing (additional information provided below) made it possible to investigate the ambient noise energy propagation direction as a potential factor. Specifically, the 2D array processing was used to investigate whether the azimuthal variability in H/V amplitude was a result of strong ambient noise energy impinging from a single dominant direction. This however was found not to be the case, as the directionality of ambient noise over the periods of interest (i.e., approximately 1 to 2 s) was found to vary substantially with time. Thus, the azimuthal differences in H/V amplitude were found not to result from strong polarization of the ambient noise. Another potential explanation for this phenomena, proposed by Matashuma et al. (\citeyear{matsushima_effect_2014}), was that azimuthal differences in H/V data, amplitude and period, may be caused by irregular subsurface topography and/or faulting. It is certainly plausible that the abruptly changing $T_{0,H/V}$ values and azimuthal irregularities in the H/V data in the northern region of Aotea Quay indicate that irregular subsurface feature(s) exists beneath this area of the port. However, determining what exactly the feature(s) may be requires further research.

\section*{Development of Shear Wave Velocity Profiles}

Vs profiles were developed for six reference locations where active- and passive-surface wave testing was performed (refer to Figures 1 and 2).  Active, MASW experiments at each of the six reference locations were used to resolve the shallow velocity structure. Passive, MAM arrays were used to characterize the deep shear wave velocity structure. The MAM arrays were deployed in seven locations (refer to Figure 2), including: two in the hydraulic fill areas of Aotea Quay (i.e., AQ and LY), four in the reclamation fill (i.e., BNZ, A2, CS, and MO), and one that stretched over both types of surficial fill (i.e., BIG). Note that the BIG array was deployed specifically to extract low frequency/long wavelength dispersion data that could be used to constrain the stiffness of the greywacke bedrock across the port. As such, it was not deployed in conjunction with smaller MAM arrays and/or active-source MASW testing. Rather, this low frequency dispersion data was used to supplement the dispersion data collected using smaller MAM arrays at the other reference locations.

MASW testing at each of the six reference locations utilized 24 vertical 4.5-Hz geophones (Geospace Technologies GS-11D). Several linear array receiver spacings ranging from 0.5-2 m were used where possible to develop broadband active experimental dispersion data. Small receiver spacings were needed to capture higher frequency/shorter wavelength data capable of resolving the stiff near-surface crust, composed of compacted fill, asphalt pavement, and subbase, present in most areas. For all arrays, Rayleigh wave content was generated by striking vertically downward directly on the pavement with a 7.3 kg sledge hammer at four distinct shot locations between 5 m and 20 m off both ends of the array. Waveforms were recorded for 1.5 s with a 0.5 s pre-trigger delay and a sampling rate of 0.5 ms. Ten successive records per shot location were recorded and stacked in the time domain to create a single record with an increased signal-to-noise ratio.

MASW time records were analyzed using several different 2D transformation methods \citep{nolet_array_1976, zywicki_advanced_1999} coupled with the multiple source-offset technique for identifying near-field contamination and quantifying dispersion uncertainty \citep{cox_surface_2011}. Dispersion data influenced by near-field effects and/or significant offline noise were eliminated. Dispersion data from each source-offset location were then used to compute mean and $\pm$ one standard deviation representative experimental dispersion data.

MAM testing involved deploying arrays of eight to ten three-component broadband seismometers and leaving them undisturbed to simultaneously collect ambient wave energy. Recording times for each array varied depending on the size of the array, but generally lasted 30 minutes to 1 hour. While it is typically preferred to use nested circular or triangular arrays, array shapes and sizes at CentrePort had to be varied based on spatial constraints at each site. The exact array configurations used at each site are visualized in Figure 2. Additional information about the MAM array configurations, including the maximum and minimum interstation distances and the theoretical array resolution limits ($k_{min}/2$) \citep{wathelet_array_2008}, are provided in Table S1 in the electronic supplement. For each array, the theoretical resolution wavelength ($\lambda_{res}$ is equal to $ 2\pi / k_{min} / 2$, and the resolution depth ($d_{res}$) is equal to $\lambda_{res}/2$  The smaller MAM arrays had $d_{res}$ values on the order of 50 m, while the larger arrays had $d_{res}$ values ranging from approximately 100 m – 385 m. However, it is important to understand that these values represent purely theoretical resolution limits, and that it is common for data of high quality to extend beyond the theoretical limits. This was certainly observed for the data acquired at CentrePort, as the dispersion data from the smaller arrays agreed well with the dispersion data from the larger arrays beyond the theoretical array resolution limits. Closely overlapping dispersion data from arrays of various sizes provides confidence for relaxing the array resolution limits in order to profile deeper when needed.

MAM time records were analyzed using the 2D high resolution frequency-wavenumber (HFK) method \citep{capon_high-resolution_1969}.  Recordings from the vertical component of each array were divided into 3 - 6 minute windows, which were processed individually. Time windows containing large oscillations, which stem from high-amplitude noise in the near-field, were eliminated. Dispersion data from the MAM arrays were used to compute mean and $\pm$ one standard deviation Rayleigh wave dispersion estimates. The experimental dispersion curves from active-source and passive-wavefield testing were then combined to create broadband, representative experimental dispersion data for each reference location. While spatial averaging within the extents of the array is inherent in all surface wave data processing, we do not believe that spatial variability had an abnormally significant impact on the dispersion data obtained from most of the MAM arrays deployed at the port.  This judgement is based on the fact that the $T_{0,H/V}$ values for any given array do not vary substantially.  Hence, while $T_{0,H/V}$ does change abruptly in some areas of the port, the surface wave arrays do not generally span across these areas. Thus, the 1D velocity models derived from inversion of the surface wave data are expected to be reasonable interpretations for the velocity structure within the bounds of each array. All inversions for this study were performed using the Dinver module in the open-source software Geopsy. The forward problem in Dinver is computed using the transfer matrix approach \citep{thomson_transmission_1950, haskell_dispersion_1953, dunkin_computation_1965, knopoff_matrix_1964} to solve for the theoretical modes of surface wave propagation associated with each trial ground model. Ground models are composed of an assumed number of layers, with each layer described by a thickness, mass density, Vs, and compression wave velocity (Vp) or Poisson’s ratio. Dinver uses a global search method (neighborhood algorithm) to locate ground models within a pre-defined parameterization that yield acceptable misfit values between the theoretical and experimental data \citep{wathelet_surface-wave_2004}.  For this work, misfit values were computed using a combined approach that considers both the goodness of fit to the experimental dispersion data and the fundamental site frequency estimated from $T_{0,H/V}$. The experimental fundamental site frequency was inferred from the average H/V spectral ratio peak ($f_{0,H/V}$) within a given array, while the theoretical fundamental site frequency was inferred from the Rayleigh wave ellipticity peak ($f_{0,R}$) calculated for a given trial ground model. Misfit values in this study were computed using Equation 2 (modified from Wathelet et al. \citeyear{wathelet_surface-wave_2004}).

\begin{equation}
    m_{d,e} = w_d m_d + w_e m_e =
    w_d \sqrt{\sum_{i=1}^{n_f}\frac{(x_{d,i}-x_{c,i})^2}{\sigma_i^2 n_f}} 
    + w_e \sqrt{\frac{(f_{0,Ell,d} - f_{0,Ell,c})^2}{\sigma^2_{f_{0,Ell,d}}}}
\end{equation}

In Equation 2, $m_{d,e}$ is the combined misfit value based on both misfit relative to dispersion data ($m_d$) and misfit relative to the Rayleigh wave ellipticity peak ($m_e$). The terms $w_d$ and $w_e$ are user-defined weighting constants for dispersion and ellipticity, respectively, which must sum to 1.0. For this study the weighting constants were set equal to 0.5. For the dispersion misfit calculations, $x_{d,i}$ represents the Rayleigh wave phase velocity of the experimental dispersion data at frequency $f_i$; $x_{c,i}$ is the calculated theoretical Rayleigh wave phase velocity for the trial ground model at frequency $f_i$; $\sigma_i$ is the standard deviation associated with the experimental dispersion data at frequency $f_i$; and $n_f$ is the number of frequency samples considered for the misfit calculation. Similarly, for the ellipticity peak misfit calculation, $f_{0,Ell,d}$ represents the Rayleigh wave ellipticity peak associated with the field data (which is assumed to coincide with the H/V peak, or $f_{0,H/V}$), $f_{0,Ell,c}$ represents the calculated theoretical Rayleigh wave ellipticity peak for the trial ground model, and $\sigma_{f_{0,Ell,d}}$ is the standard deviation associated with the experimental ellipticity peak (which is assumed to be equal to the standard deviation of $f_{0,H/V}$, or $\sigma_{f_{0,H/V}}$. Additional details regarding calculation of the combined dispersion and fundamental site frequency misfit values are provided in Teague et al. (\citeyear{teague_development_2017}).

\begin{figure}[!]
    \centering
	\includegraphics[width=0.5\textwidth]{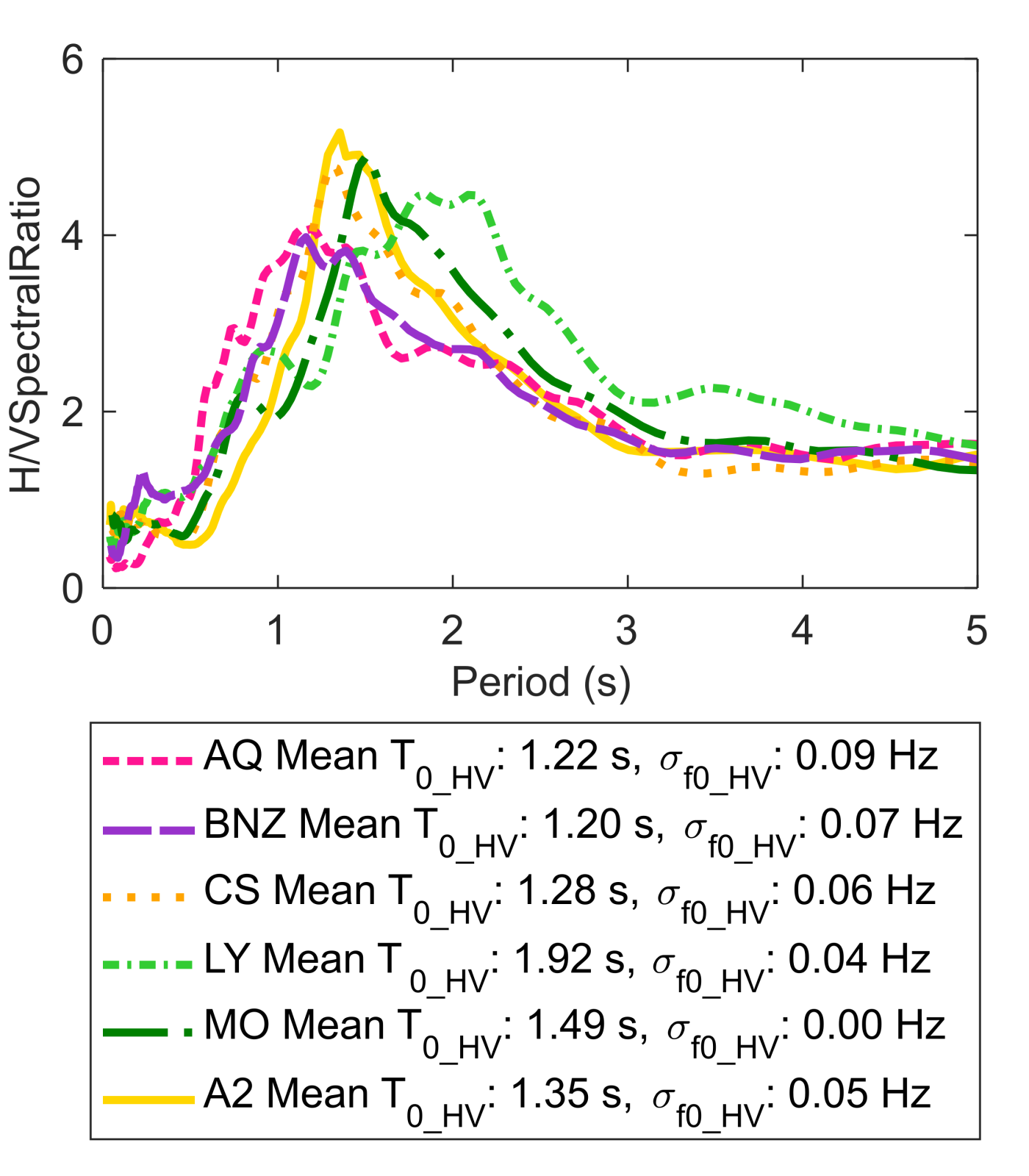}
	\caption{Mean H/V spectral ratio data for each MAM array used to develop Vs profiles at six reference locations across CentrePort.}
	\label{fig:5}
\end{figure}

The average experimental H/V curves corresponding to all six reference locations are shown in Figure \ref{fig:5}. The average H/V spectral ratio curve for each array was calculated from those stations whose peaks were considered to be “clear” based on SESAME (\citeyear{sesame_guidelines_2004}), as well as any station that exhibited a peak judged to be of good quality, though not strictly “clear”. The fundamental frequency peak (based on maximum mean amplitude) for each station composing a given array was used to calculate a mean and standard deviation to constrain the Rayleigh wave ellipticity in the inversion process for the array. For those sites where the clarity of the H/V data was considered to be less distinct (i.e., AQ, LY), inversions were performed both with and without the Rayleigh wave ellipticity peak to check the sensitivity of the resulting velocity profiles. In general, it was found that even without including the ellipticity peak as a target (i.e., using dispersion data only) the coincidental fit between $f_{0,R}$ and $f_{0,H/V}$ was found to be quite good. Therefore, as the peak in the H/V data contains an additional piece of information to help constrain the inversion, $f_{0,H/V}$ was utilized in all inversions to better resolve the depth to bedrock.

The inverse problem is known to be inherently ill-posed, nonlinear, and mix-determined, without a unique solution \citep{foti_non-uniqueness_2009, di_giulio_exploring_2012}. To address the poorly constrained nature of the inverse problem, borehole (Tonkin \& Taylor Ltd., \citeyear{tonkin__taylor_ltd_thorndon_2012}) and CPT \citep{cubrinovski_liquefaction-induced_2018} data were used to inform the inversion parameterization (i.e., number and thickness of model layers) and limit solution non-uniqueness, as recommended by Teague et al. (\citeyear{teague_development_2017}). However, as borehole/CPT information at the port was only available down to about 20 m, the layering ratio approach \citep{cox_layering_2016}, which provides a systematic method of varying the inversion parametrization in order to account for epistemic uncertainty, was utilized at greater depths where site specific information was unavailable. For each site, six layering ratios ($\Xi$) (1.3, 1.5, 2.0, 3.0, 5.0, and 7.0) were investigated to address non-uniqueness in the solution of the inverse problem. This approach resulted in trial layered earth models with as many as 28 layers (corresponding to $\Xi$=1.3) and as few as 8 layers (corresponding to $\Xi$=7.0).  Inversions were run using a minimum of 500,000 trial earth models for high layering ratios with few layers, and with as many as 1.4 million trial earth models for low layering ratios with many layers.

\section*{Results for the Thorndon Wharf (A2) Reference Location}

A detailed discussion of the results for a single representative reference location will be presented here, followed by the presentation of the results for all reference locations in the form of calculated depths to soft (Vs $>$ 760 m/s) and hard (Vs $>$ 1500 m/s) rock, as defined by the NEHRP Site Class B and Site Class A boundaries, respectively (International Code Council, \citeyear{international_code_council_international_2015}; ASCE \citeyear{ asce_minimum_2010}). However, for completeness, dispersion data and inversion results for all six reference locations have been included in an electronic supplement.

\begin{figure}[!]
    \centering
	\includegraphics[width=\textwidth]{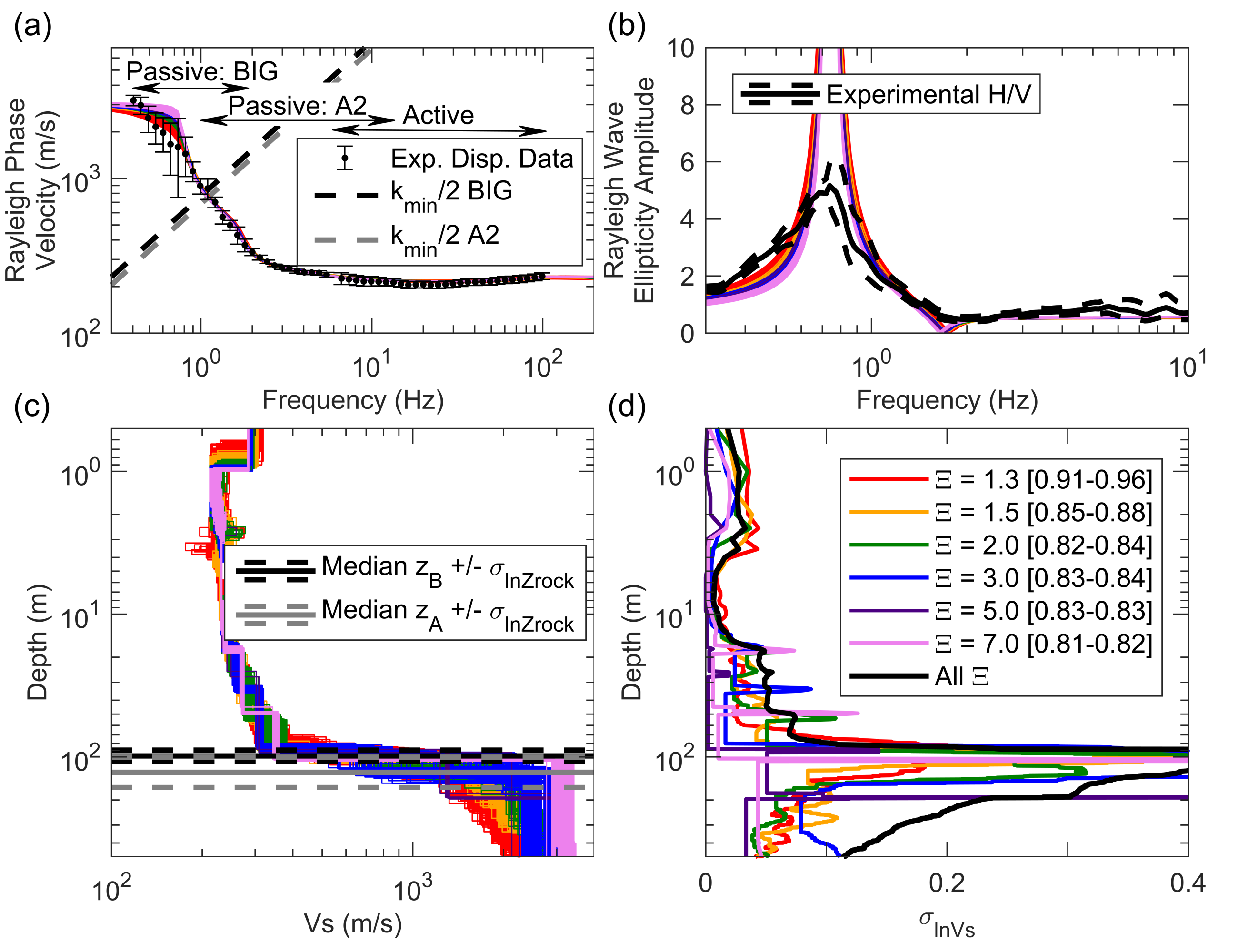}
	\caption{Inversion results for the Thorndon Wharf (A2) reference location. Shown for each acceptable layering ratio ($\Xi$) inversion parameterization are the 100 lowest misfit: (a) theoretical fundamental mode Rayleigh wave dispersion curves with the experimental dispersion data and the theoretical array resolution limits ($k_{min}/2$); (b) theoretical Rayleigh wave ellipticity with the lognormal median and $\pm$ one standard deviation experimental H/V data; (c) Vs profiles shown to depths of 500 m with the lognormal median and $\pm$ one standard deviation depth to soft and hard rock ($Z_{B}$ and $Z_{A}$ respectively); and (d) standard deviation of the natural logarithm of Vs ($\sigma_{ln,Vs}$). Note that the range of misfit values associated with each suite of velocity profiles are provided inside the brackets of the figure legend located in panel (d).}
	\label{fig:6}
\end{figure}

The Thorndon Wharf site (A2), is located in the southeast corner of CentrePort (refer to Figure 1). MAM arrays used at A2 included three nested, non-concentric circles with diameters of 50, 150, and 300 m (refer to Figure 2). MASW active-source testing involved two lines of geophones, one at 1 m spacing and a second at 2 m spacing. The dispersion data extracted from these arrays is shown in Figure \ref{fig:6}a. The active and passive dispersion data form a broadband experimental dispersion curve from approximately 0.5 – 100 Hz. Note that the dispersion data below approximately 1.0 Hz was obtained exclusively from the BIG array. While this dispersion data exceeds the array resolution limits ($k_{min}/2$), and is therefore more uncertain, it can still be used to help better constrain the velocity of the bedrock.

The 600 lowest misfit theoretical dispersion curves (i.e., 100 lowest misfit models for each of the six trial layering ratio inversion parameterizations) obtained from inversions at location A2 are shown in comparison to the experimental dispersion data in Figure 6a. The misfit values for all 600 models (shown in brackets in the figure's legend) ranged from 0.81 to 0.96 across all layering ratios. When considering the meaning of dispersion misfit values, it is important to remember that they can only be used to compare the relative quality of fit for dispersion curves from the same site, and cannot be used to compare the quality of fit from one site to another due to variable complexity in data from site-to-site \citep{griffiths_surface-wave_2016}.  However, a dispersion misfit value less than 1.0 essentially means that, on average across the frequency range of the experimental data, the theoretical model falls within the ±1 standard deviation bounds of the experimental data \citep{cox_layering_2016}.

The Rayleigh wave ellipticity curves for the 600 lowest misfit models are shown in comparison to the experimental H/V data in Figure 6b. They are observed to match the fundamental frequency inferred from the average H/V curve very well. Note that the relative amplitudes of the ellipticity and H/V curves are meaningless, only the relative locations of the peaks are important. The Vs profiles for the 600 lowest misfit models are shown in Figure 6c. The Vs profiles from all layering ratio parameterizations agree quite well, with the biggest differences observed for the depth and stiffness of bedrock. From these velocity profiles, the median depth to soft rock ($Z_{B}$ = depth where Vs $>$ 760 m/s) and hard rock ($Z_{A}$ = depth where Vs $>$ 1500 m/s) $\pm$ one lognormal standard deviation have been calculated. These median depths are $Z_{B}$ = 98 m and $Z_{A}$ = 128 m, which are significantly less than the $d_{res}$ values for the largest circular A2 array and the BIG array, approximately 340 and 390 m respectively (refer to Table S1).

Intra- and inter-inversion lognormal standard deviations in Vs ($\sigma_{ln,Vs}$) for the 100 lowest misfit models associated with each layering ratio parameterization are shown in Figure 6d. The inter-inversion variability is clearly higher than the intra-inversion variability, illustrating the importance of considering multiple inversion parameterizations when attempting to realistically quantify Vs uncertainty associated with surface wave inversion. Note that “spikes” in the $\sigma_{ln,Vs}$ values do not represent uncertainty in Vs, but uncertainties in the locations of boundaries between layers. On average, the $\sigma_{ln,Vs}$ values for the soil deposits at A2 are less than 0.1. The $\sigma_{ln,Vs}$ for bedrock is closer to 0.15.

\begin{figure}[!]
    \centering
	\includegraphics[width=\textwidth]{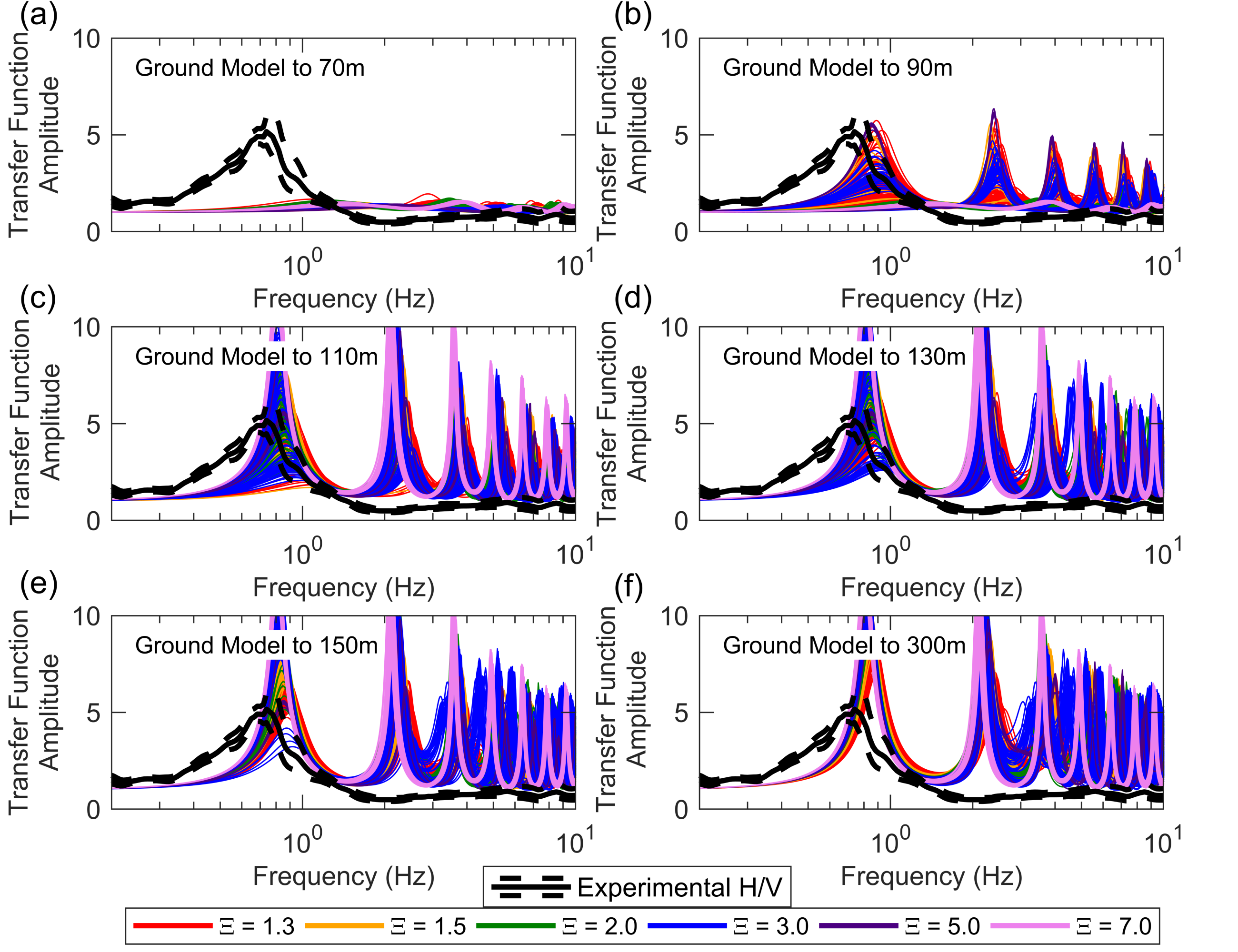}
	\caption{ Comparison of the theoretical shear wave transfer functions for the 600 lowest misfit velocity models (100 velocity models per layering ratio inversion parameterization) at the Thorndon Wharf (A2) reference location truncated at depths of: (a) 70 m, (b) 90 m, (c) 110 m, (d) 130 m, (e) 150 m, and (f) 300 m.}
	\label{fig:7}
\end{figure}

To illustrate how $T_{0,H/V}$ (i.e., $1/f_{0,H/V}$) is related to the layer boundaries indicated in the Vs profiles, theoretical shear wave transfer functions were calculated from the Vs profiles after truncating them at several different depths. The linear viscoelastic transfer functions were computed for “outcrop” conditions using damped, horizontal layers over an elastic halfspace \citep{kramer_geotechnical_1996}.  Small-strain damping ratios for soil layers were based on Darendeli (\citeyear{darendeli_development_2001}) and set equal to 0.5\% for rock.  The theoretical shear wave transfer functions for the 600 lowest misfit Vs profiles truncated at six depths are shown in Figure \ref{fig:7}. As the profiles are extended to greater depths, moving from Figure 7a to 7f, the lowest frequency peak from the linear elastic shear wave transfer function ($f_{0,S}$) is shown to approach $f_{0,H/V}$. At a depth of 70 m (Figure 7a), there is no significant contrast in the velocity profiles and therefore no significant peak in the shear wave transfer function is observed. As the depth increases from 70 m to 90 m (Figure 7b), it is clear that some of the velocity profiles begin to have a $f_{0,S}$ that are closer to $f_{0,H/V}$. Note that the additional higher frequency peaks in the transfer function are, in this case, representative of higher modes which are not expected to be represented in the H/V data. When the depth is increased to 110 m (Figure 7c), nearly all profiles show $f_{0,S}$ at a frequency slightly higher than $f_{0,H/V}$, although some of the peaks are quite subdued. As the depth is further increased (Figures 7d, 7e, and 7f) all profiles show a clear and strong $f_{0,S}$ at a frequency slightly higher than $f_{0,H/V}$. In the case of the A2 reference location, the $f_{0,S}$ for most velocity profiles is apparently dictated by velocity contrasts between 130 and 150 m. For A2, it is noted that this depth range exists just below the median depth to hard rock ($Z_A$ = 128 m), as estimated from the velocity profiles in Figure 6c. It is also interesting to observe that $f_{0,S}$ does not perfectly match $f_{0,H/V}$ even though $f_{0,R}$ does (refer to Figure 6d).

\section*{Estimates of the Depth to Rock Across CentrePort}

\begin{figure}[!]
    \centering
	\includegraphics[width=0.6\textwidth]{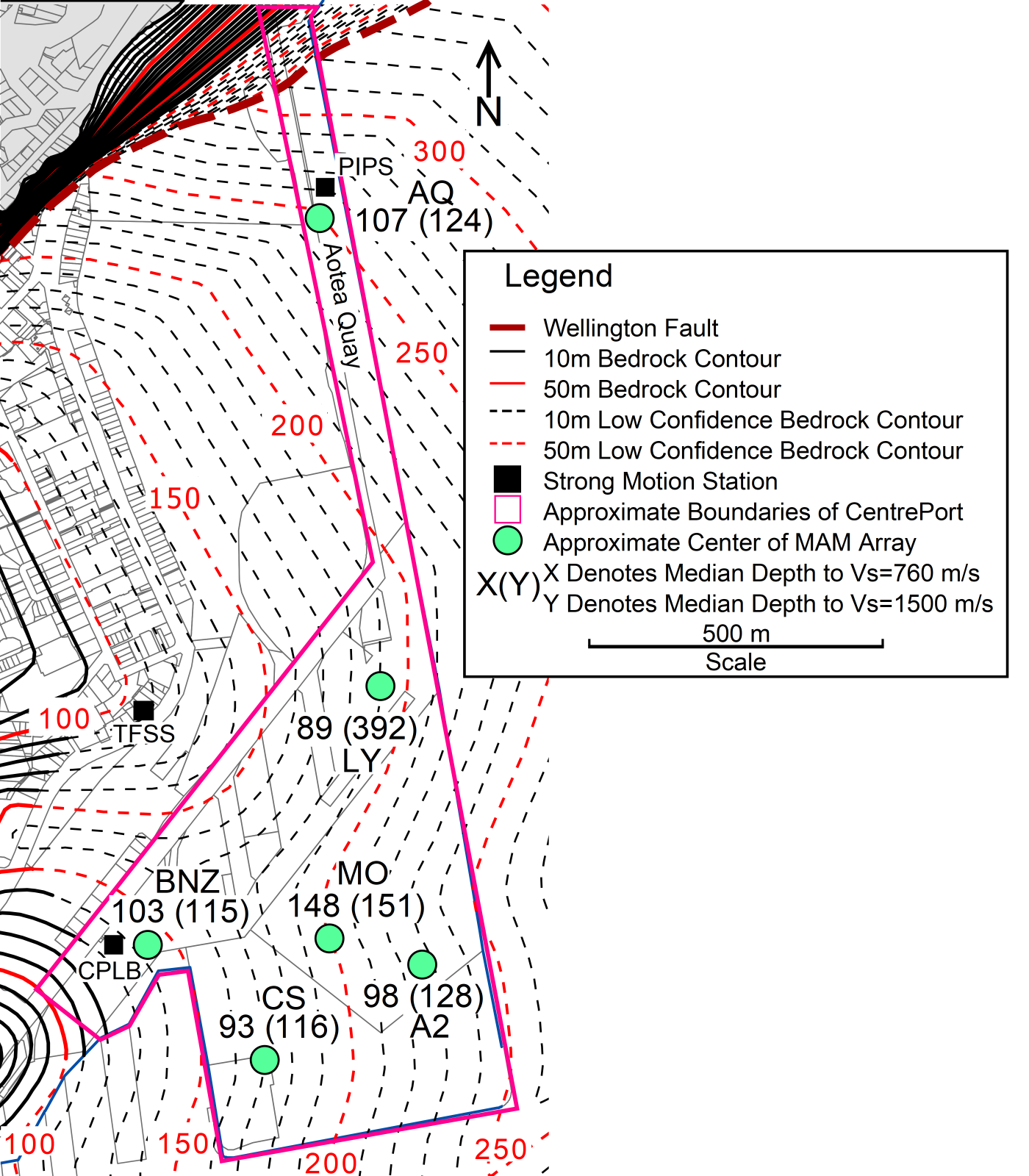}
	\caption{Median depths to soft and hard rock (Vs $>$ 760 m/s and Vs $>$ 1500 m/s, respectively) as determined from Vs profiles developed from surface wave testing at six reference locations across CentrePort. The depths to soft and hard rock are shown in meters as X(Y), where X is the median depth to soft rock and Y is the median depth to hard rock. These values are shown in comparison with bedrock contours estimated by Semmens et al. (\citeyear{semmens_its_2010}).}
	\label{fig:8}
\end{figure}

The median depths to soft and hard rock for all six reference locations are summarized in Figure \ref{fig:8}. At least two things are notable from these depth to rock estimates. First, $Z_B$ only varies from approximately 90-150 m across the entire port. These values are significantly less than the depth to bedrock values estimated by Semmens et al. (\citeyear{semmens_its_2010}), which range from approximately 150-300 m, as indicated by the depth to rock contour lines in Figure 8. But recall that the Semmens et al. (\citeyear{semmens_its_2010}) depth to bedrock estimates in CentrePort were based on extrapolation from measurements located hundreds of meters from the port’s boundaries. While extrapolations were informed by surface topography and regional expertise, it is important to understand that they have been indicated with low confidence and cannot be taken as “ground truth”. Second, a pattern similar to that observed for A2, where the depths to soft and hard rock are relatively close to one another (i.e., within 30 m), was observed for five of the six reference locations across the port. However, the depths to rock at the Log Yard reference location do not follow this pattern. In fact the depth to hard rock at the LY site is inferred to be over 300 m deeper than the depth to soft rock, which is unexpected based on current geologic knowledge and deserves further consideration.

\section*{Further Consideration for the Log Yard (LY) Reference Location}

\begin{figure}[!]
    \centering
	\includegraphics[width=\textwidth]{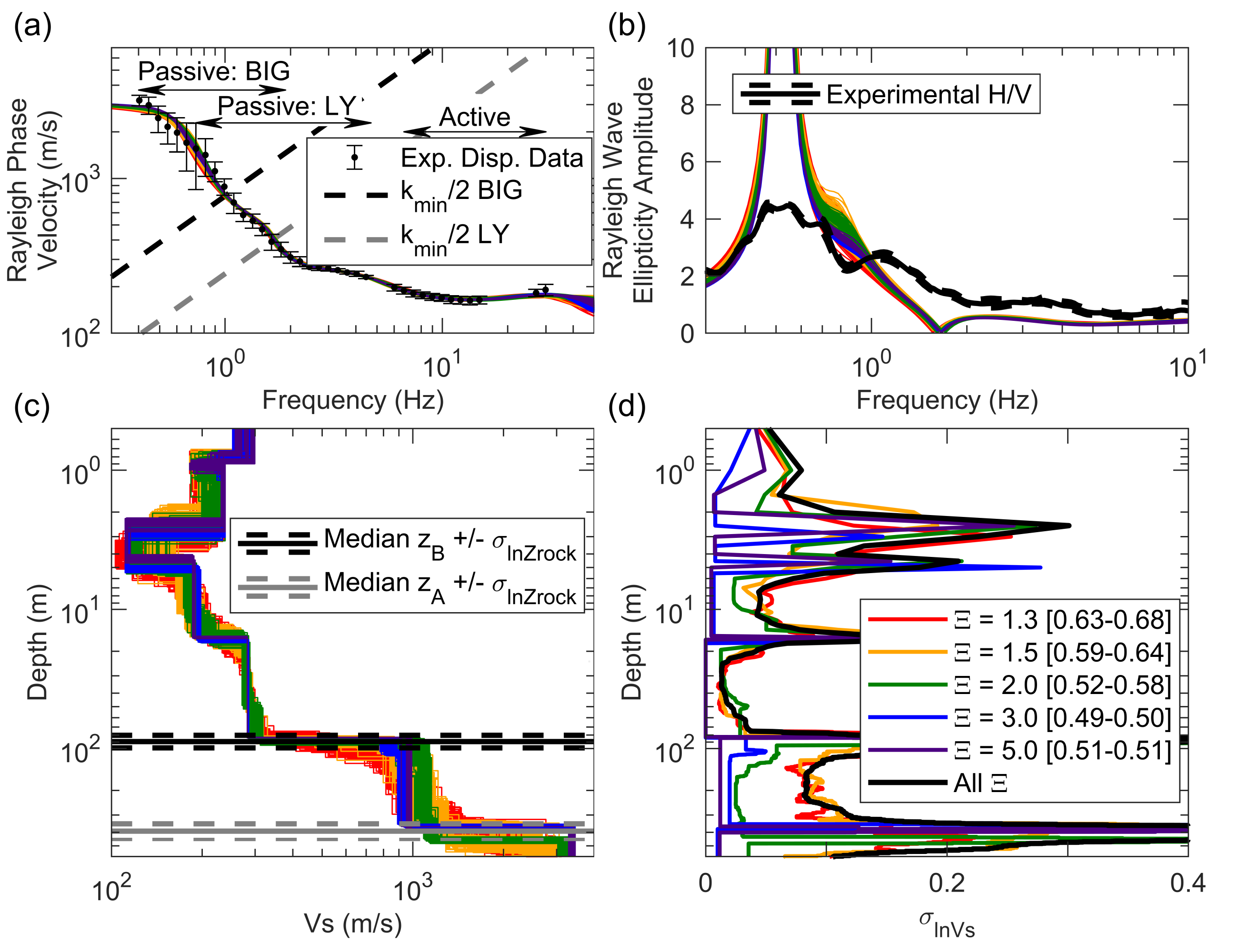}
	\caption{Inversion results for the Log Yard (LY) reference location Shown for each acceptable layering ratio ($\Xi$) inversion parameterization are the 100 lowest misfit: (a) theoretical fundamental mode Rayleigh wave dispersion curves with the experimental dispersion data and the theoretical array resolution limits ($k_{min}/2$); (b) theoretical Rayleigh wave ellipticity with the lognormal median and $\pm$ one standard deviation experimental H/V curve; (c) Vs profiles shown to depths of 600 m with the lognormal median and $\pm$ one standard deviation depth to soft and hard rock ($Z_B$ and $Z_A$ respectively); and (d) standard deviation of the natural logarithm of Vs ($\sigma_{ln,Vs}$). Note that the range of misfit values associated with each suite of velocity profiles are provided inside the brackets of the figure legend located in panel (d).}
	\label{fig:9}
\end{figure}

Passive-wavefield MAM testing at the Log Yard reference location involved a large T-shaped array with one of the legs extending 135 m and the other 75 m. The active-source MASW testing involved a 46 m array with 2 m geophone spacing. As shown in Figure \ref{fig:9}a, the experimental dispersion data extracted from these arrays was of high quality and aligned well with the low frequency dispersion data extracted from the BIG array down to about 0.7 Hz, which is well below the theoretical array resolution limits of either array. Thus, while the theoretical resolution depth of the LY array (refer to Table S1) is only about 120 m, we feel confident that the dispersion data from the LY array has a much greater resolution depth. This presumption is based on the fact that the low frequency data is very distinct and in good agreement with the data from the BIG array, which has a theoretical resolution depth of approximately 385 m.

The theoretical dispersion curves obtained from the 100 lowest misfit models for each layering ratio inversion are compared with the experimental dispersion data in Figure 9a. All theoretical dispersion curves resulted in a dispersion misfit values between 0.49-0.68. Furthermore, the $f_{0,R}$ shown in Figure 9b agree well with $f_{0,H/V}$ for the LY site. However, it should be noted that the peak in the average experimental H/V data for LY is more broad than the peaks in the average H/V data for other arrays (refer to Figure 5) and does not give what could be confidently considered as a “clear” peak according to the SESAME (\citeyear{sesame_guidelines_2004}) guidelines. Nonetheless, it is evident that the H/V data indicates a longer fundamental site period than the other array locations, supporting a deeper bedrock contrast and/or softer overlying sediments. The Vs profiles shown in Figure 9c indicate both of these characteristics in comparison to the Vs profiles for A2 (refer to Figure 6c). Specifically, the near-surface materials at LY are much softer than those for A2, and while the median depth to soft rock is slightly shallower at LY, the median depth to hard rock is hundreds of meters deeper.

The apparent significant depth to hard rock beneath LY was re-investigated through various trial inversions, including consideration of different modal interpretations of the dispersion data, with and without using the H/V peak, and with and without using the supplementary lowest frequency data from the BIG array. None of these alternate inversions changed the resulting depth to hard rock significantly. Thus, we have to consider that this apparently anomalous result is a viable solution. While purely speculation, it is possible that the large incised river channel/valley extending beneath the port in the vicinity of the seismic station TFSS (refer to Figure 1) is related to the drastically differing depths to soft and hard rock at the LY reference location.

\begin{figure}[!]
    \centering
	\includegraphics[width=\textwidth]{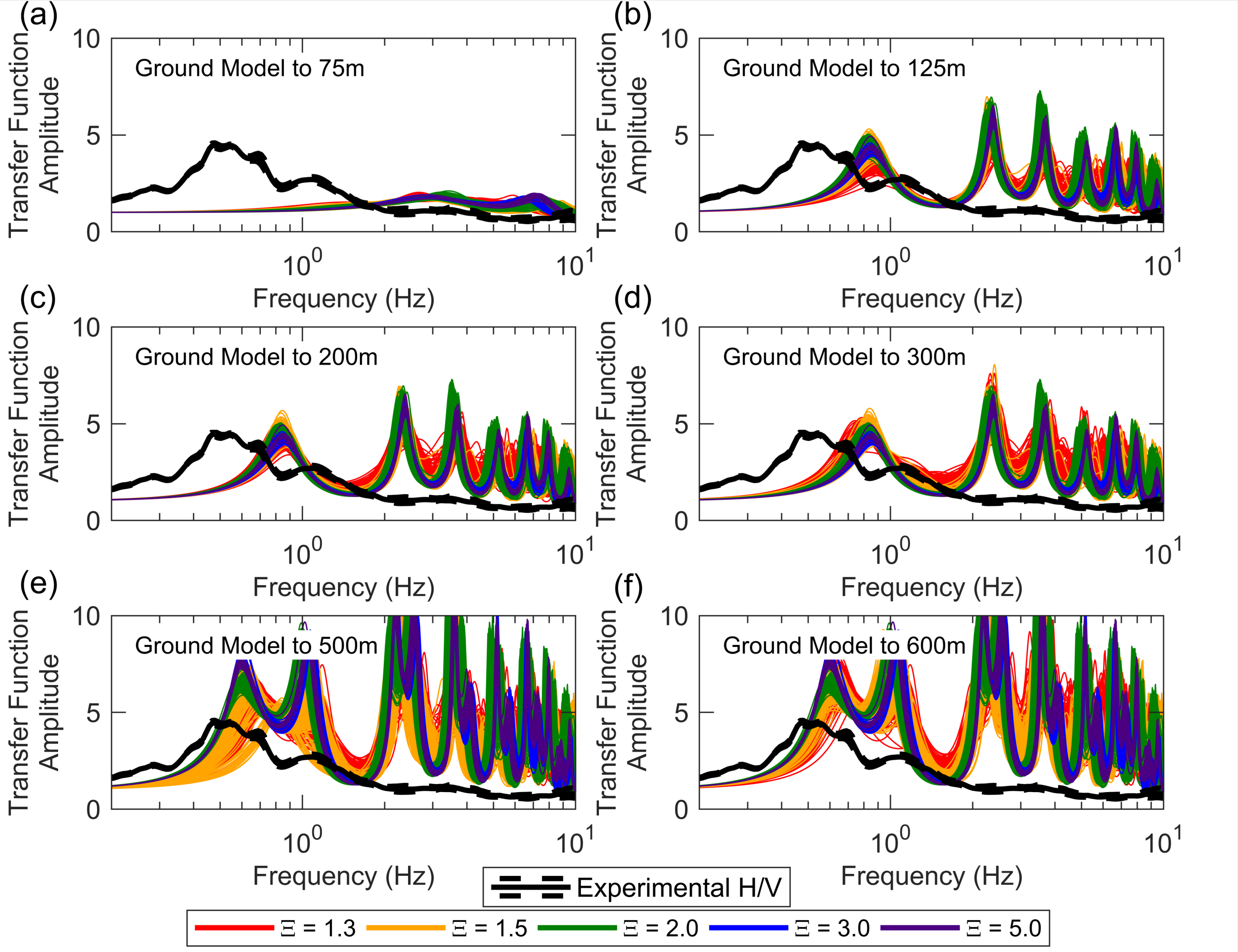}
	\caption{Comparison of the theoretical shear wave transfer function for the 500 lowest misfit velocity models (100 velocity models per layering ratio inversion parameterization) at the Log Yard (LY) reference location truncated at depths of: (a) 75 m, (b) 125 m, (c) 200 m, (d) 300 m, (e) 500 m, and (f) 600 m.}
	\label{fig:10}
\end{figure}

Regardless of the root cause driving differences in the inferred depths to soft and hard rock, it is important to investigate which impedance contrasts are most important to modeling site response.  To illustrate how $T_{0,H/V}$ is related to the layer boundaries indicated in the LY Vs profiles, theoretical shear wave transfer functions were calculated from the Vs profiles truncated at various depths. These results are shown in Figure \ref{fig:10}. Vs profiles truncated above the median depth to soft rock (refer to Figure 10a) do not show a strong $f_{0,S}$. Profiles truncated at depths below the median soft rock contact, but above the hard rock contact, do show a strong peak (refer to Figures 10b, 10c, and 10d), but it is not aligned with $f_{0,H/V}$, indicating that that Vs profiles truncated at these depths are not capturing a key feature of the site signature and should not be used in site response analyses \citep{teague_site_2016, teague_measured_2017}. Profiles truncated below the median depth to the hard rock contact (refer to Figure 10e and 10f) have clear $f_{0,S}$ peaks that reasonably match both low frequency peaks in the experimental H/V curve. Therefore, in the case of LY it is necessary to extend the velocity profiles to depths below the hard rock contact in order to capture the fundamental site frequency inferred from H/V.  Once again, $f_{0,S}$ does not perfectly match $f_{0,H/V}$ for any of the Vs profiles even though $f_{0,R}$ does (refer to Figure 8d).

\section*{Dominant Site Periods from Spectral Amplifications at Ground Motion Stations near CentrePort}

\begin{figure}[!]
    \centering
	\includegraphics[width=\textwidth]{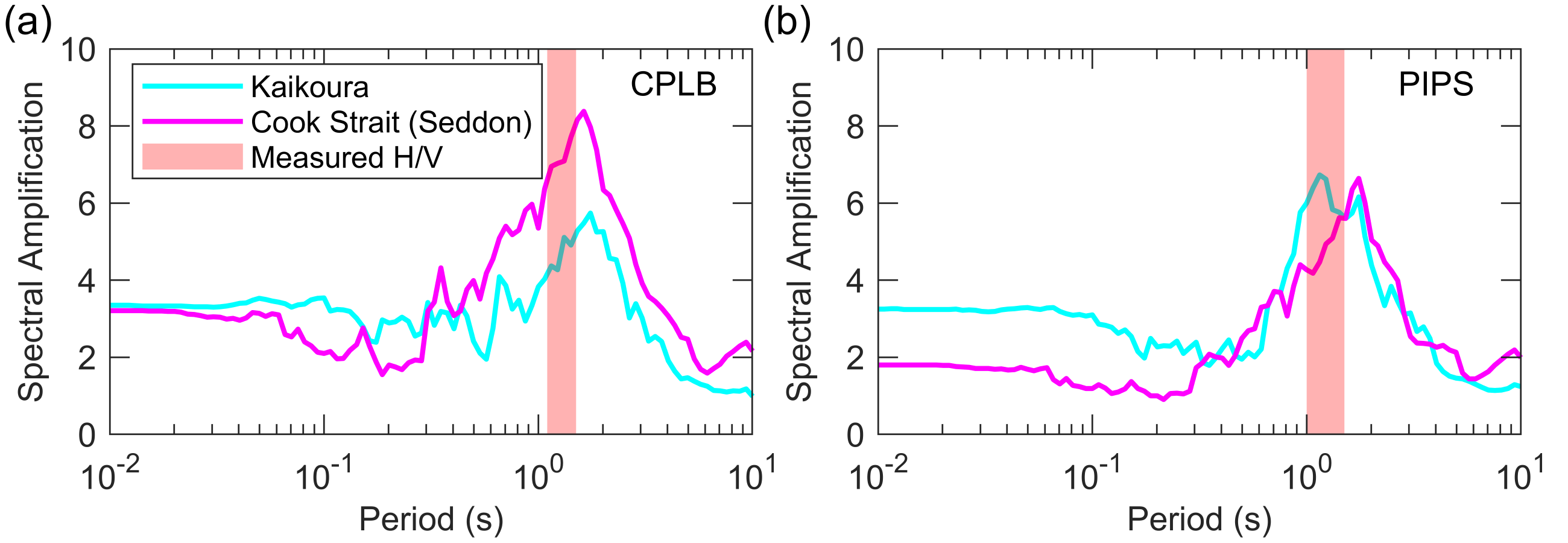}
	\caption{Spectral amplification ($SA_{site} / SA_{reference}$) for the (a) CPLB and (b) PIPS ground motion stations calculated relative to the nearby rock reference station POTS. Ground motions recorded during the 2016 Kaik\=oura and 2013 Cook Strait earthquakes are shown relative to the ranges in ambient vibration H/V site period ($T_{0,H/V}$) measured near the CPLB and PIPS stations.}
	\label{fig:11}
\end{figure}

The peaks in the H/V spectral ratio measurements across the port (refer to Figure 3) represent estimates of the small-strain, linear viscoelastic fundamental site period at that location. Therefore, local amplification of earthquake ground motions should occur for periods at or above these linear viscoelastic estimates of $T_0$, depending on the degree of nonlinearity/softening induced by larger strain ground motions. Site amplifications during previous earthquakes can be estimated using the ratio of the pseudo-acceleration response spectra for the location of interest (i.e., the amplified ground motion) and a reference rock station (i.e., the input ground motion). Spectral amplifications for the CPLB and PIPS strong motion stations located within CentrePort have been calculated relative to the nearby POTS reference rock station by Bradley et al. (\citeyear{bradley_ground_2017}) for both the 2016 Kaik\=oura earthquake and the 2013 Cook Strait earthquake. These spectral amplification functions are shown in Figure \ref{fig:11}a and 11b for the CPLB and PIPS ground motion stations, respectively. Also shown in these figures are the ranges in $T_{0,H/V}$ values derived from our H/V measurements taken closest to the ground motion stations. The peak in the spectral amplification functions for both CPLB and PIPS agree favorably with $T_{0,H/V}$ values obtained from ambient vibrations. While the $T_{0,H/V}$ values slightly under estimate the period of absolute maximum spectral amplification, this is to be expected for several potential reasons, including site period elongation due to nonlinear soil behavior associated with higher intensity earthquake ground motions and other complicating factors such as 3D basin edge effects. These issues will require future research efforts to fully understand and the $T_{0,H/V}$ estimates and deep Vs profiles presented herein will provide key inputs into subsequent back- and forward-site response analyses.

\section*{Conclusions}

Dynamic site characterization studies at CentrePort using H/V spectral ratio measurements as well as active-source and passive-wavefield surface wave testing (i.e., MASW and MAM) allowed for a detailed study of the spatial variation of the fundamental site period,  shear stiffness, and depths to soft (Vs $>$ 760 m/s) and hard (Vs $>$ 1500 m/s) rock across the port. $T_{0,H/V}$ at CentrePort was found to generally increase from south to north from approximately 1.0 to 2.2 seconds. However, two notable and abrupt discontinuities were observed: one in the vicinity of the Log Yard reference location, which was driven by changes in both near-surface soil stiffness and depth to hard rock, and a second in the vicinity of northern Aotea Quay, which was likely driven only by changes in the depth to hard rock and complex subsurface 3D velocity structure, as indicated by azimuthal variations in the H/V spectra. Linear viscoelastic shear wave transfer functions determined from the Vs profiles developed at several reference locations were truncated at various depths to investigate which impedance contrasts need to be modeled for capturing site response at the port. These investigations indicated that Vs profiles need to be extended down to hard rock in order for the transfer functions to capture $T_{0,H/V}$. The bedrock depths beneath CentrePort previously estimated by Semmens et al. (\citeyear{semmens_its_2010}) were found, in most cases, to significantly overestimate the depth to bedrock. $T_{0,H/V}$ measurements made in the vicinity of the CPLB and PIPS stations proved to be fairly good indicators of the predominant periods of spectral amplification recorded at these stations during both the 2016 Kaik\=oura earthquake and the 2013 Cook Strait earthquake. However, future research efforts utilizing the $T_{0,H/V}$ estimates and deep Vs profiles presented herein will be required to capture important ground motion characteristics caused by soil nonlinearity and 3D basin edge effects. While this paper has focused exclusively on the dynamic site characterization of CentrePort, our team has also collected a great deal of data outside of the port in greater Wellington. Ultimately, all of this data will need to be synthesized to understand complex patterns of ground motion amplification across the city caused by rapidly varying 3D surface and subsurface topography coupled with soft soil conditions. 

\section*{Data and Resources}

Seismograph time records used in this study were acquired from GeoNet through the New Zealand Strong Motion Database \url{www.geonet.org.nz/data/supplementary/nzsmdb}. The database can be accessed by \url{ftp://ftp.geonet.org.nz/strong/processed/Proc/nzsmb/} (last accessed August 2017). Metadata for each strong motion station can be accessed through \url{https://magma.geonet.org.nz/delta/app} (last accessed September 2017). The open-source software Geopsy is available from www.geopsy.org (last accessed November 2016). Inversion analyses were performed on the Texas Advanced Computing Center (TACC) resources Stampede and Stampede2.

\section*{Acknowledgements}

This work was supported by U.S. National Science Foundation (NSF) grant CMMI-1724915. Financial support was also provided by the New Zealand Earthquake Commission (EQC) under the Capability Building Fund at the University of Auckland, through the Natural Hazards Research Platform (NHRP) grant “Kaik\=oura Earthquake response – geotechnical characterization of CentrePort reclamations” through MBIE, and QuakeCoRE through Technology Platform 2. This is QuakeCoRE publication number 0216.  However, any opinions, findings, conclusions, or recommendations expressed in this paper are those of the authors and do not necessarily reflect the views of either NSF or EQC. We would also like to acknowledge the collaboration efforts of Tiffany Krall and many other members of the CentrePort Ltd.\ team for allowing access to the port and proving assistance with our dynamic site characterization efforts. We would like to thank the anonymous reviewers and editors for their efforts to help strengthen this work.

\bibliographystyle{plainnat}
\bibliography{centreport}

\section*{Appendix: Electronic Supplement}

This electronic supplement contains additional information regarding the 2D MAM surface wave arrays deployed at CentrePort, Wellington, New Zealand for shear wave velocity profiling. The information provided in Table \ref{table:S1} for each array includes: shape, number of stations, minimum and maximum interstation distance, theoretical array resolution limit, and approximate theoretical resolution depth. Following Table S1 are six figures (Figures \ref{fig:S1}, \ref{fig:S2}, \ref{fig:S3}, \ref{fig:S4}, \ref{fig:S5}, and \ref{fig:S6}) that document the dispersion data and inversion results for all six reference locations used to map the depth to bedrock across CentrePort. Note that the inversion results from the Thorndon Warf (A2) and Log Yard (LY) reference locations are discussed at length in the main body of the article and are only repeated here for completeness. The reader is referred to the main body of the article for additional information that will facilitate understanding of these figures.

\begin{landscape}

\setcounter{table}{0}
\renewcommand{\thetable}{S\arabic{table}} 

\begin{table}[]
\centering
\caption{Array geometry and theoretical resolution limits associated with the 2D MAM arrays used for surface wave testing at CentrePort, Wellington, New Zealand.}
\begin{tabular}{@{}ccccccccc@{}}
\toprule
\begin{tabular}[c]{@{}c@{}}Reference\\ Location\end{tabular} &
  \begin{tabular}[c]{@{}c@{}}Array\\ Designation\end{tabular} &
  \begin{tabular}[c]{@{}c@{}}Array\\ Shape\end{tabular} &
  \begin{tabular}[c]{@{}c@{}}Number \\ of  Stations\end{tabular} &
  \begin{tabular}[c]{@{}c@{}}Maximum\\ Interstation\\ Distance\\ (m)\end{tabular} &
  \begin{tabular}[c]{@{}c@{}}Minimum\\ Interstation\\ Distance\\ (m)\end{tabular} &
  \begin{tabular}[c]{@{}c@{}}Theoretical\\ Array\\ Resolution\\ Limit,\\ $k_{min}/2$\\ (1/m)\end{tabular} &
  \begin{tabular}[c]{@{}c@{}}Resolution\\ Wavelength,\\ $\lambda_{res}$\\ (m)\end{tabular} &
  \begin{tabular}[c]{@{}c@{}}Resolution \\ Depth,\\ $d_{res}$ \\ (m)\end{tabular} \\ \midrule
\begin{tabular}[c]{@{}c@{}}Aotea Quay\\ (AQ)\end{tabular}                     & AQ\_T\_100 & Triangles & 10 & 100  & 10  & 0.0666 & 94  & 47  \\
\begin{tabular}[c]{@{}c@{}}BNZ Building\\ (BNZ)\end{tabular}                  & BNZ\_L\_60 & L-shape   & 10 & 60   & 5   & 0.0663 & 95  & 47  \\
\begin{tabular}[c]{@{}c@{}}Cold Store\\ (CS)\end{tabular}                     & CS\_L\_160 & L-shape   & 10 & 160  & 5   & 0.0323 & 194 & 97  \\
\begin{tabular}[c]{@{}c@{}}Log Yard\\ (LY)\end{tabular}                       & LY\_T\_135 & T-shape   & 10 & 135  & 15  & 0.0259 & 242 & 121 \\
\begin{tabular}[c]{@{}c@{}}Main Office\\ (MO)\end{tabular}                    & MO\_L\_90  & L-shape   & 10 & 90   & 5   & 0.0545 & 115 & 58  \\
\multirow{3}{*}{\begin{tabular}[c]{@{}c@{}}Thorndon Warf\\ (A2)\end{tabular}} & A2\_C\_300 & Circular  & 10 & 490  & 45  & 0.0092 & 684 & 342 \\
                                                                              & A2\_C\_150 & Circular  & 8  & 180  & 65  & 0.0162 & 389 & 194 \\
                                                                              & A2\_C\_050 & Circular  & 10 & 50   & 10  & 0.0519 & 121 & 61  \\
\begin{tabular}[c]{@{}c@{}}Big Array\\ (BIG)\end{tabular}                     & BIG\_L     & L-shape   & 10 & 1575 & 130 & 0.0082 & 769 & 385 \\ \bottomrule
\end{tabular}
\label{table:S1}
\end{table}

\end{landscape}

\setcounter{figure}{0}
\renewcommand{\thefigure}{S\arabic{figure}} 

\begin{figure}[!]
    \centering
	\includegraphics[width=\textwidth]{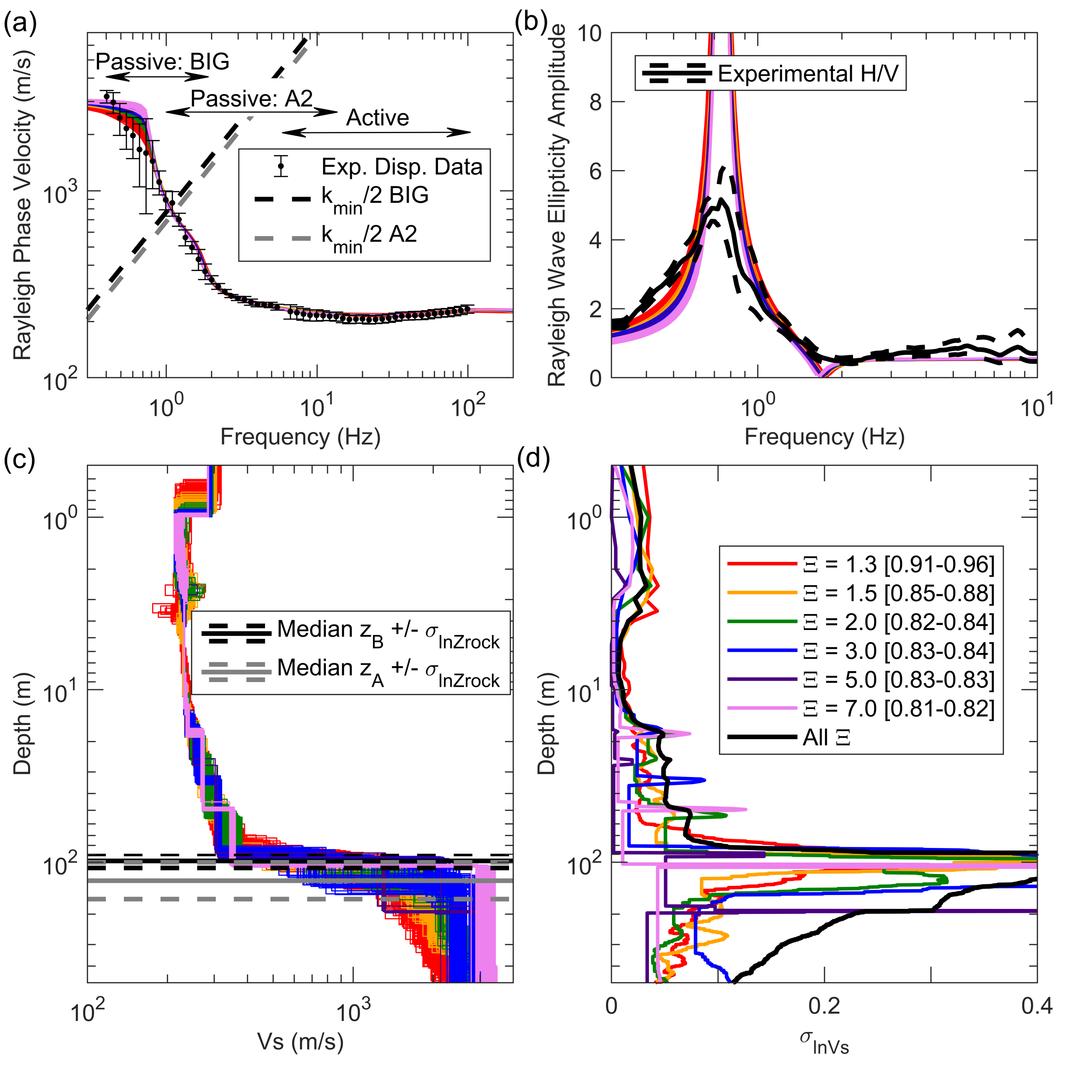}
	\caption{Inversion results for the Thorndon Wharf (A2) reference location. Shown for each acceptable layering ratio ($\Xi$) inversion parameterization are the 100 lowest misfit: (a) theoretical fundamental mode Rayleigh wave dispersion curves with the experimental dispersion data and the theoretical array resolution limits ($k_{min}/2$); (b) theoretical Rayleigh wave ellipticity with the lognormal median and $\pm$ one standard deviation experimental H/V data; (c) Vs profiles shown to depths of 500 m with the lognormal median and $\pm$ one standard deviation depth to soft and hard rock ($Z_B$ and $Z_A$ respectively); and (d) standard deviation of the natural logarithm of Vs ($\sigma_{ln,Vs}$). Note that the range of misfit values associated with each suite of velocity profiles are provided inside the brackets of the figure legend located in panel (d).}
	\label{fig:S1}
\end{figure}

\begin{figure}[!]
    \centering
	\includegraphics[width=\textwidth]{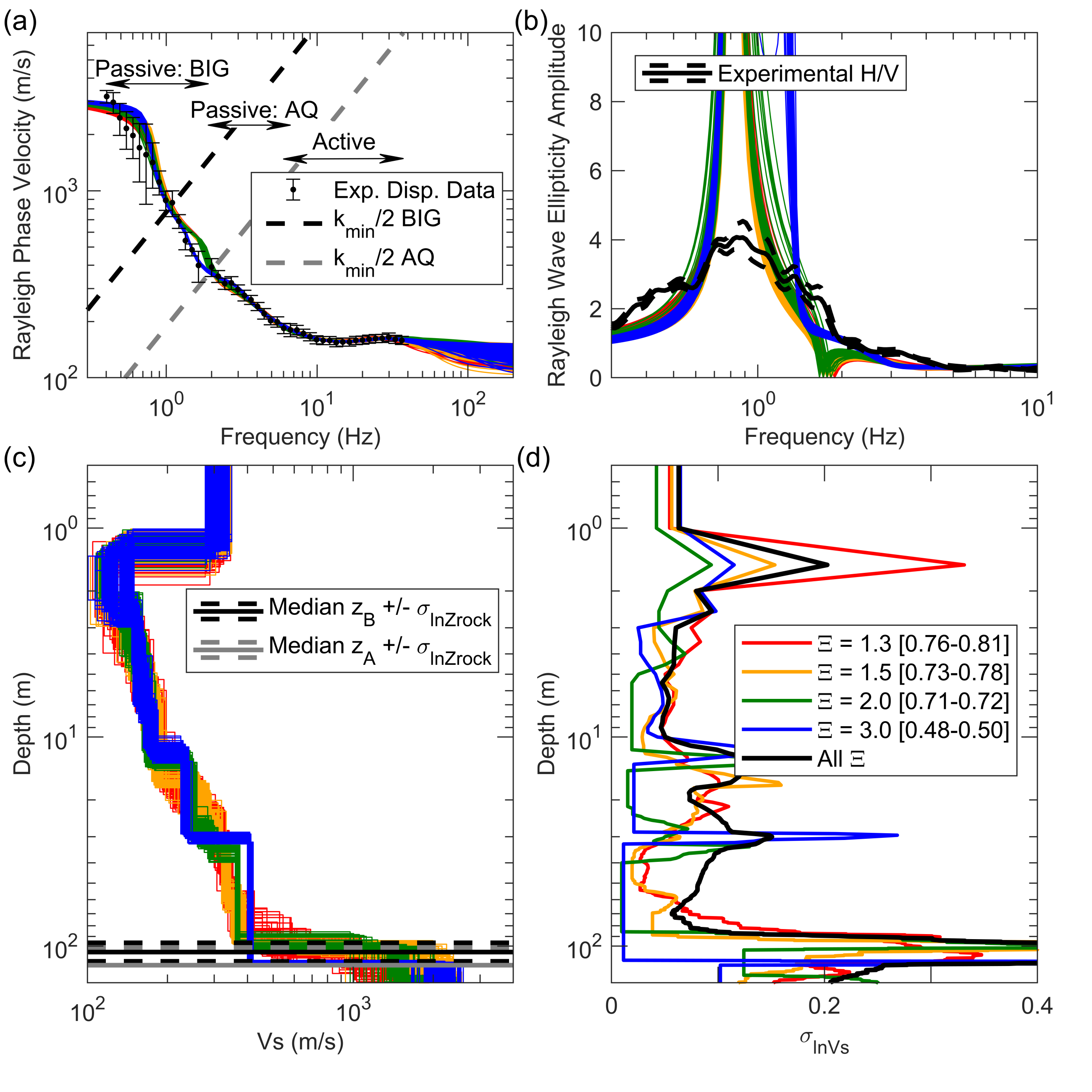}
	\caption{Inversion results for the Aotea Quay (AQ) reference location. Shown for each acceptable layering ratio ($\Xi$) inversion parameterization are the 100 lowest misfit: (a) theoretical fundamental mode Rayleigh wave dispersion curves with the experimental dispersion data and the theoretical array resolution limits ($k_{min}/2$); (b) theoretical Rayleigh wave ellipticity with the lognormal median and $\pm$ one standard deviation experimental H/V data; (c) Vs profiles shown to depths of 150 m with the lognormal median and $\pm$ one standard deviation depth to soft and hard rock ($Z_B$ and $Z_A$ respectively); and (d) standard deviation of the natural logarithm of Vs ($\sigma_{ln,Vs}$). Note that the range of misfit values associated with each suite of velocity profiles are provided inside the brackets of the figure legend located in panel (d).}
	\label{fig:S2}
\end{figure}

\begin{figure}[!]
    \centering
	\includegraphics[width=\textwidth]{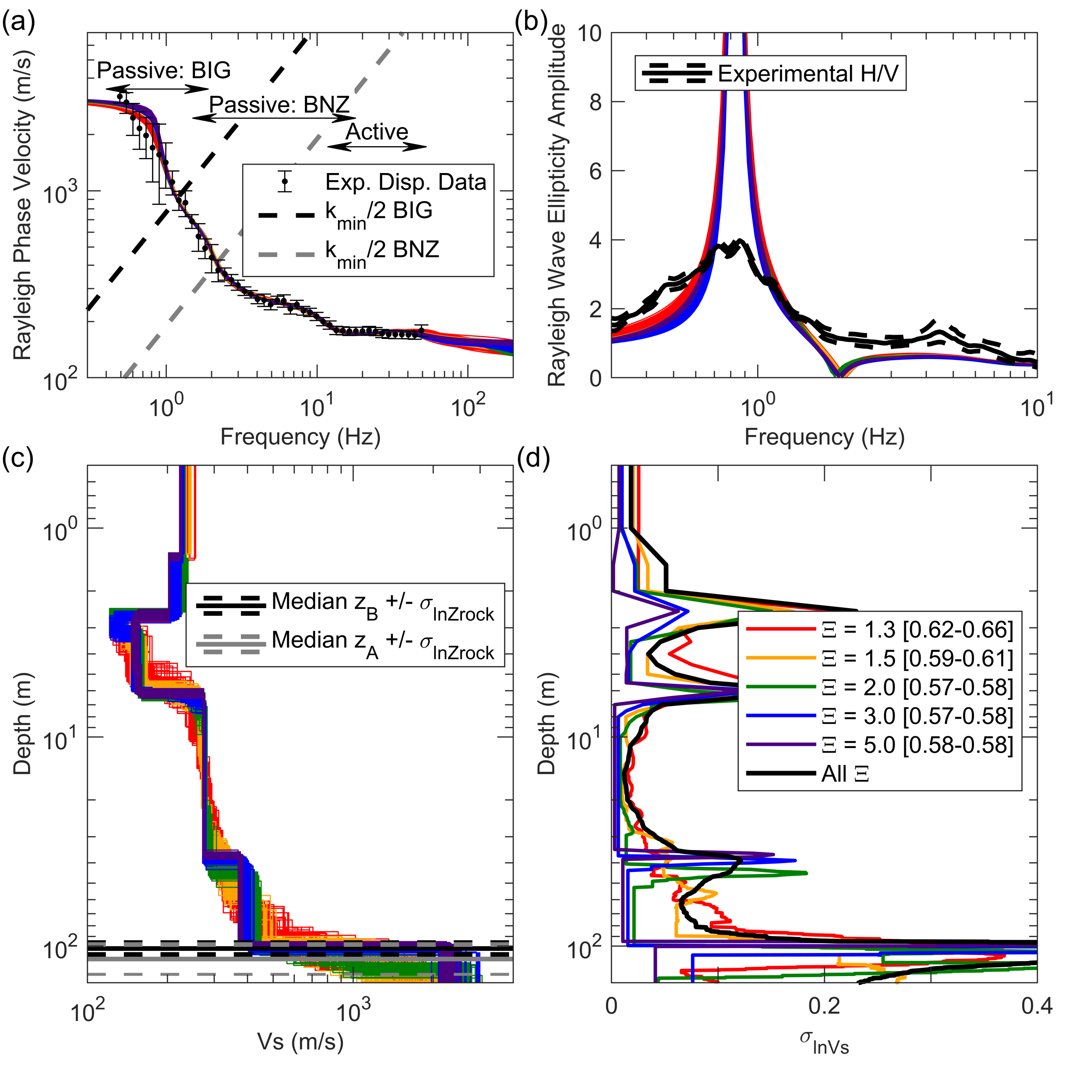}
	\caption{Inversion results for the BNZ Building (BNZ) reference location. Shown for each acceptable layering ratio ($\Xi$) inversion parameterization are the 100 lowest misfit: (a) theoretical fundamental mode Rayleigh wave dispersion curves with the experimental dispersion data and the theoretical array resolution limits ($k_{min}/2$); (b) theoretical Rayleigh wave ellipticity with the lognormal median and $\pm$ one standard deviation experimental H/V data; (c) Vs profiles shown to depths of 150 m with the lognormal median and $\pm$ one standard deviation depth to soft and hard rock ($Z_B$ and $Z_A$ respectively); and (d) standard deviation of the natural logarithm of Vs ($\sigma_{ln,Vs}$). Note that the range of misfit values associated with each suite of velocity profiles are provided inside the brackets of the figure legend located in panel (d).}
	\label{fig:S3}
\end{figure}

\begin{figure}[!]
    \centering
	\includegraphics[width=\textwidth]{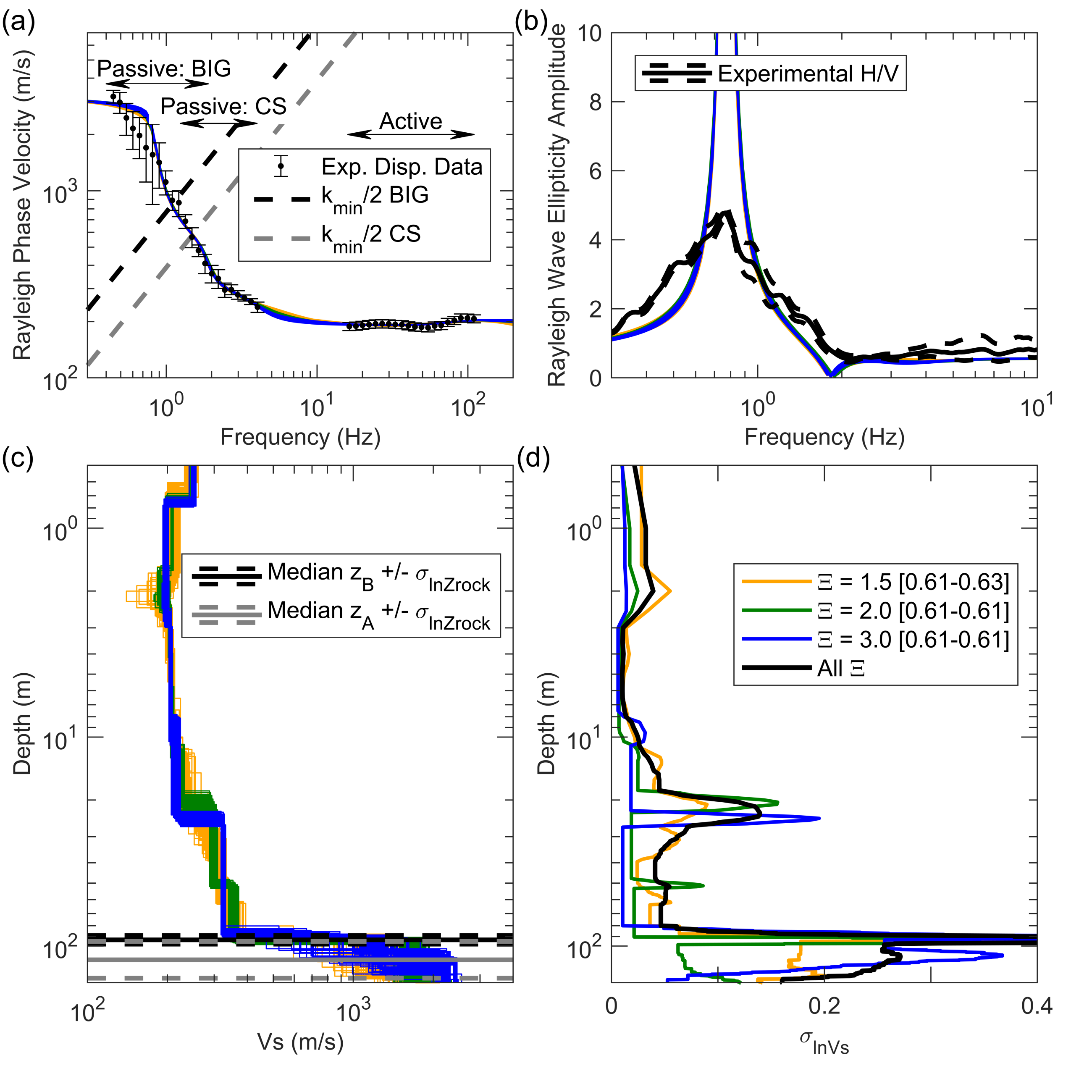}
	\caption{Inversion results for the Cold Store (CS) reference location. Shown for each acceptable layering ratio ($\Xi$) inversion parameterization are the 100 lowest misfit: (a) theoretical fundamental mode Rayleigh wave dispersion curves with the experimental dispersion data and the theoretical array resolution limits ($k_{min}/2$); (b) theoretical Rayleigh wave ellipticity with the lognormal median and $\pm$ one standard deviation experimental H/V data; (c) Vs profiles shown to depths of 150 m with the lognormal median and $\pm$ one standard deviation depth to soft and hard rock ($Z_B$ and $Z_A$ respectively); and (d) standard deviation of the natural logarithm of Vs ($\sigma_{ln,Vs}$). Note that the range of misfit values associated with each suite of velocity profiles are provided inside the brackets of the figure legend located in panel (d).}
	\label{fig:S4}
\end{figure}

\begin{figure}[!]
    \centering
	\includegraphics[width=\textwidth]{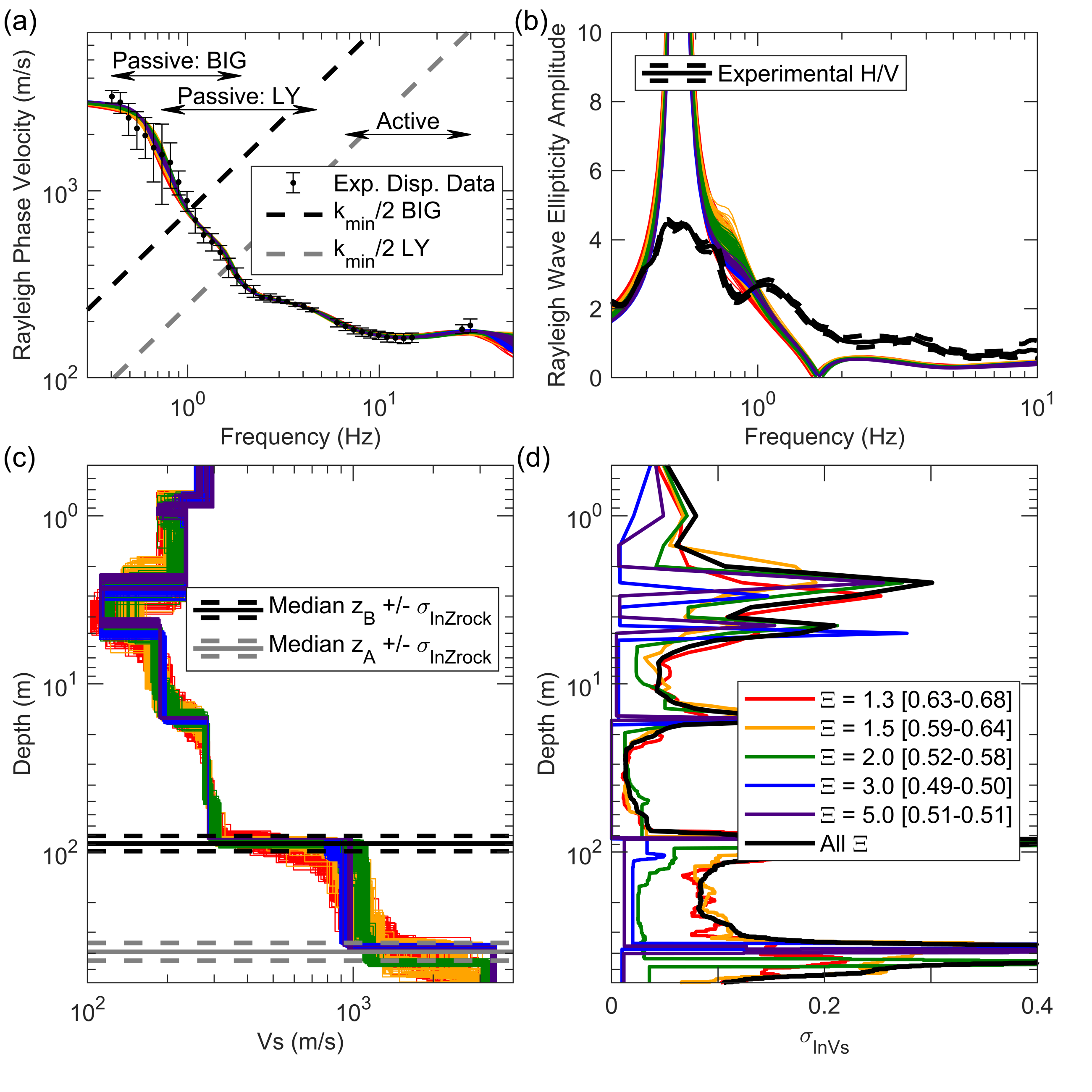}
	\caption{Inversion results for the Log Yard (LY) reference location. Shown for each acceptable layering ratio ($\Xi$) inversion parameterization are the 100 lowest misfit: (a) theoretical fundamental mode Rayleigh wave dispersion curves with the experimental dispersion data and the theoretical array resolution limits ($k_{min}/2$); (b) theoretical Rayleigh wave ellipticity with the lognormal median and $\pm$ one standard deviation experimental H/V curve; (c) Vs profiles shown to depths of 600 m with the lognormal median and $\pm$ one standard deviation depth to soft and hard rock ($Z_B$ and $Z_A$ respectively); and (d) standard deviation of the natural logarithm of Vs ($\sigma_{ln,Vs}$). Note that the range of misfit values associated with each suite of velocity profiles are provided inside the brackets of the figure legend located in panel (d).}
	\label{fig:S5}
\end{figure}

\begin{figure}[!]
    \centering
	\includegraphics[width=\textwidth]{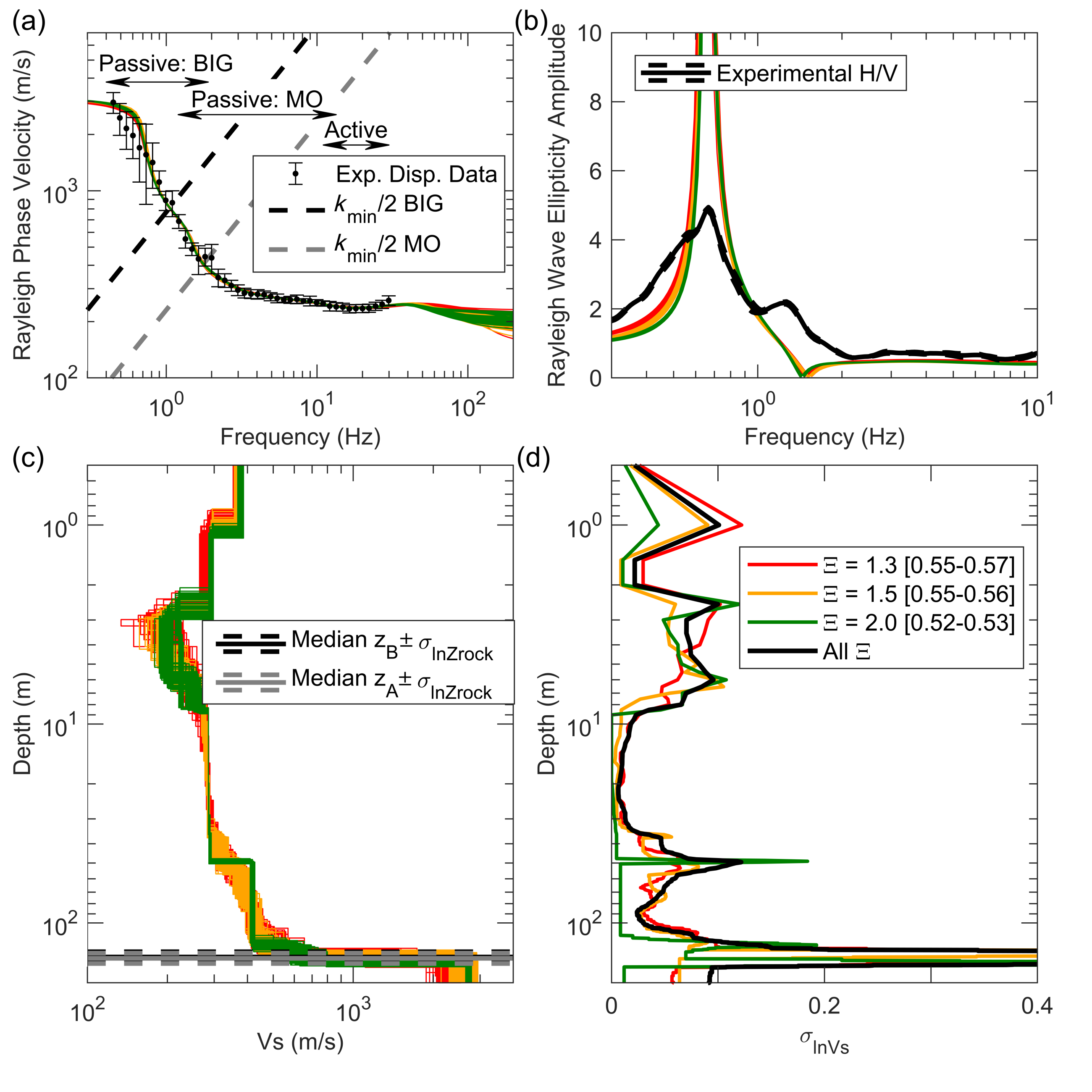}
	\caption{Inversion results for the Main Office (MO) reference location. Shown for each acceptable layering ratio ($\Xi$) inversion parameterization are the 100 lowest misfit: (a) theoretical fundamental mode Rayleigh wave dispersion curves with the experimental dispersion data and the theoretical array resolution limits ($k_{min}/2$); (b) theoretical Rayleigh wave ellipticity with the lognormal median and $\pm$ one standard deviation experimental H/V data; (c) Vs profiles shown to depths of 150 m with the lognormal median and $\pm$ one standard deviation depth to soft and hard rock ($Z_B$ and $Z_A$ respectively); and (d) standard deviation of the natural logarithm of Vs ($\sigma_{ln,Vs}$). Note that the range of misfit values associated with each suite of velocity profiles are provided inside the brackets of the figure legend located in panel (d).}
	\label{fig:S6}
\end{figure}

\end{document}